\documentclass[printer]{aa}  

\usepackage{graphicx}
\usepackage{amsmath}
\usepackage{bm}
\newcommand{\Rm}{\mathrm{Rm}}

\newcommand{\LA}{\Lambda_\mathrm{A}}
\newcommand{\OK}{\Omega_\mathrm{K}}
\newcommand{\Macc}{\dot{M}_\mathrm{acc}}

\usepackage{txfonts}
\usepackage[urlcolor=blue]{hyperref}
\begin{document}

   \title{A systematic description of wind-driven protoplanetary discs\thanks{The self-similar solutions presented in this paper are available for download on github: \href{https://github.com/glesur/PPDwind}{https://github.com/glesur/PPDwind}}}

   \author{Geoffroy R. J. Lesur
          \inst{1}          }

   \institute{Univ. Grenoble Alpes, CNRS, IPAG, 38000 Grenoble, France\\
              \email{geoffroy.lesur@univ-grenoble-alpes.fr}}

   \date{Received --------; accepted January 24, 2021}

 
  \abstract
   {}
   {Planet forming discs are believed to be very weakly turbulent in the regions outside of 1 AU. For this reason, it is now believed that magnetised winds could be the dominant mechanism driving accretion in these systems. However, there is today no self-consistent way to describe discs subject to a magnetised wind, in a way similar to the $\alpha$ disc model. In this article, I explore in a systematic way the parameter space of wind-driven protoplanetary discs and present scaling laws which can be used in reduced models \emph{à la} $\alpha$ disc. }
   {I compute a series of self-similar wind solutions, assuming the disc is dominated by ambipolar and Ohmic diffusions. These solution are obtained by looking for stationary solutions in the finite-volume code PLUTO using a relaxation method and continuation. }
   {Self similar solutions are obtained for values of plasma $\beta$ ranging from $10^2$ to $10^{8}$, for several Ohmic and ambipolar diffusion strengths. Mass accretions rates of the order of $10^{-8}\,M_\odot/\mathrm{yr}$ are obtained for poloidal field strength $\beta=O(10^4)$ or equivalently $1\,\mathrm{mG}$ at 10 AU. In addition, the ejection efficiency is always close to 1, implying that wind mass loss rate can be larger than the inner mass accretion rate if the wind-emitting region is large. The resulting magnetic lever arms are typically lower than 2, possibly reaching 1.5 in weakest field cases. Remarkably, the mean transport properties (accretion rate, mass loss rate) depend mostly on the field strength and much less on the disc diffusivities or surface density. The disc internal structure is nevertheless strongly affected by Ohmic resistivity, strongly resistive discs being subject to accretion at the surface while ambipolar only models lead to mid-plane accretion. Finally, I provide a complete set of scaling laws and semi-analytical wind solutions, which can be used to fit and interpret observations. }
   {Magnetised winds are unavoidable in protoplanetary discs as soon as they are embedded in an ambient poloidal magnetic field. Very detailed disc microphysics are not always needed to describe them, and simplified models such as self-similar solutions manage to capture most of the physics seen in full 3D simulations. The remaining difficulty to get a complete theory of wind-driven accretion lies in the transport of the large scale field, which remains poorly constrained and not well understood.}


   \maketitle
%

\section{Introduction}

Protoplanetary discs are relatively cold and dense objects, which typically last for a few million years around young stellar objects (YSOs). However, it is well known that these discs are primarily \emph{accretion} discs, in which matter is slowly falling onto the central star. This accretion rate has now been measured in dozens of objects, and typically lies in the range of $10^{-10}$ to $10^{-7}\,M_\odot/\mathrm{yr}$. Accretion in astrophysical discs is usually thought to be due to a magnetohydrodynamic (MHD) instability, the magneto-rotational instability \citep[MRI,][]{Balbus91}, which transports angular momentum outwards and mass inwards. It was quickly realised that applying the MRI in the context of protoplanetary discs is a difficult task because of the dramatically low ionisation fraction expected in these objects. Indeed, the inclusion of Ohmic \citep{Gammie96} and ambipolar diffusions \citep{Perez11} led to the conclusion that MRI cannot account for the observed accretion rates in these discs by two to three orders of magnitude.

The lack of a proper mechanism to trigger turbulence and angular momentum transport in protoplanetary discs revived an old idea mostly abandoned since the finding of the MRI: magnetised winds. Indeed, while the disc bulk might be too diffusive to sustain MHD turbulence, it is still weakly coupled to the ambient large scale field originating from the parent cloud of the YSO. This ambient field can in principle be enough to trigger a magnetic breaking of the disc, leading to mass accretion and the formation of a wind. This idea was originally proposed in the context of YSOs by \cite{Wardle93}. It was later revived via numerical simulations, first in local shearing box setups \citep{Bai13a,Simon13,lesur14} and then in global simulations \citep{Gressel15,Bethune17,Bai17,Wang19,Gressel20}. All of these simulations include Ohmic and ambipolar diffusion, while some of them also include the Hall effect which is believed to be important in the densest regions of the disc (around 1 AU). The strength of these diffusions is in turn computed from ionisation and chemical models of various complexity. In addition, almost all of these models include some sort of heating of the disc atmosphere, designed to mimic non-thermal radiation heating expected in these systems \citep[e.g.][]{Thi19}. In the end of the day, each published model is of very high complexity, and only a handful of simulations can be performed exploring a (very) limited subspace of parameter space.

In this context, the purpose of this work is to take a step back, simplifying the physics in order to explore in a more or less systematic manner the coupling between protoplanetary discs and magnetised winds. This work follows in essence the approach initiated by \cite{Ferreira97} on globally self-similar models and later extended to viscous \citep{Casse00} and weakly magnetised \citep{Jacquemin19} discs. I however use a relatively different technic to get the wind solutions, which I present in the next section, along with the physical model. I then focus on a fiducial series of wind solution and explore their physical properties. Finally, I vary the disc diffusivities to get an overview of the more exotic configurations before concluding.
   
\section{Model}
In the following, I consider a magnetised disc orbiting a central object of mass $M$. The disc is subject to the gravitational pull of the central object only (self-gravity being neglected), and the Lorentz force due to electrical currents. Disc is assumed to be weakly ionised, hence I consider a generalised Ohm's law which include Ohmic and ambipolar diffusivities. These diffusivities are prescribed in section \ref{sec:diffusivities}.

\subsection{Equations}
In the following, I will use either spherical $(r,\theta,\varphi)$ coordinates or cylindrical $(R,\varphi,z)$ coordinates, depending on the context. Note that the cylindrical radius is written in upper case, while the spherical radius is lower case. I solve the non-ideal MHD equations, here in spherical coordinates:
\begin{align}
\label{eq:cont}\partial_t\rho+\bm{\nabla\cdot }\rho\bm{v}& =0 ,\\
\label{eq:motion}\partial_t\rho\bm{v}+\bm{\nabla\cdot}\rho\bm{vv}&= - \bm{\nabla}\rho c_s^2+\frac{\bm{J\times B}}{c} -\frac{GM\rho}{r^2}\bm{e}_r ,\\
\label{eq:induct}\partial_t\bm{B}&=-\bm{\nabla\times \mathcal{E}}\\
\label{eq:Ohm}\bm{\mathcal{E}}=-\bm{v\times B}+\frac{4\pi}{c}\Big( \eta_{\rm O}\bm{J}&+\eta_{\rm H}\bm{J\times \hat{B}}-\eta_{\rm A}\bm{J\times \hat{B}\times \hat{B}}\Big)
\end{align}
where I have defined the isothermal sound speed $c_s(\bm{r})$, assuming the flow was locally isothermal, the plasma current $\bm{J}=c\bm{\nabla\times B}/4\pi$, the gravitational constant $G$ the magnetic field direction $\bm{\hat{B}}$, the electromotive field $\bm{\mathcal{E}}$ and the Ohmic, Hall and ambipolar diffusivities $\eta_{\rm O}$, $\eta_{\rm H}$ and $\eta_{\rm A}$. Note that the diffusivities are also functions of space and magnetic field strength. I stress that \emph{no turbulence is assumed} in this model, only molecular magnetic diffusivities which results from the low ionisation fraction of the plasma. If accretion is occurring, it is only the result of torques which are self-consistently computed in the model.

\subsection{Self-similar ansatz\label{sec:ss}}
In the context of protoplanetary discs, the strong diffusivities are known to almost suppress all non-axisymmetric structures \citep{Bethune17,Bai17} in the regions outside 1 AU. For that reason, I assume that the flow is 2.5D: I conserve three components for $\bm{v}$ and $\bm{u}$ but neglect $\varphi$ derivatives.

In order to simplify the problem even more, I will assume the flow is globally self-similar. This implies that as one moves away from the central object, the flow "looks" the same. This approach has several advantages: first it avoids issues with the inner boundary conditions which are usually problematic in global numerical models. Second, it allows us to explore systematically the parameter space at exquisite resolution with a limited numerical cost, as the problem becomes essentially a 1D problem.  I follow \cite{Ferreira93} and define the self-similar scaling for any field $Q$ as
\begin{align}
\label{eq:ssDefinition}
Q(r,\theta)=\Bigg(\frac{r}{r_0}\Bigg)^{\gamma_Q}\tilde{Q}(\theta),	
\end{align}
where $\gamma_Q$ is the self-similar exponent and $\tilde{Q}$ is a 1D function which completely determines the flow. The scaling of the gravitational force as $1/r^2$ imposes the self-similar scaling to all of the other components, i.e
\begin{align*}
\gamma_v=-\frac{1}{2}\ ;\ 	\gamma_B=-\frac{5}{4}\ ;\  \gamma_\rho=-\frac{3}{2}\ ;\  \gamma_\eta=\frac{1}{2}.
\end{align*}
Similarly to the velocity field, the sound speed $c_s$ is proportional to $r^{-1/2}$, and $\tilde{c_s}$ is a prescribed function of $\theta$. Note also that the self-similar scaling is here written in spherical coordinates, but can be transformed into cylindrical ($R,z,\varphi$) coordinates by noting that $r=R\sin(\theta)$ and $z/R=\tan^{-1}(\theta)$:
\begin{align*}
Q(R,z)=\Bigg(\frac{R}{R_0}\Bigg)^{\gamma_Q}	\hat{Q}(z/R)
\end{align*}
 where $\hat{Q}(z/R)=\sin^{\gamma_Q}(\theta)\tilde{Q}(\theta)$.

In the following, it will be useful to define the Keplerian velocity $v_\mathrm{K}(R)\equiv\sqrt{GM/R}$ and the Keplerian angular frequency $\OK\equiv v_\mathrm{K}/R$. I will also use the disc geometrical thickness $h\equiv c_s(R,z=0)/\OK(R)$. Note that by construction from the self-similar scaling, $h/R\equiv\varepsilon $ is a constant of the problem.

\subsection{Diffusivities\label{sec:diffusivities}}

The diffusivities $\eta$ are usually computed from complex thermo-chemical network \citep[e.g.][]{Thi19}. It is customary to use dimensionless numbers to quantify the diffusivities: the Ohmic Reynolds number $\Rm$ and the ambipolar Elsasser number $\LA$ defined as
\begin{align}
\Rm&\equiv \frac{\Omega_K h^2}{\eta_\mathrm{O}}\\
\LA&\equiv\frac{V_\mathrm{A}^2}{\Omega_K\eta_\mathrm{A}}
\end{align}
these dimensionless number are more useful than the more traditionally used Elsasser numbers since in most cases they do not depend on the magnetic field strength \citep{Wardle99}. Hence, they only depend on the gas properties (density, temperature, composition).

The use of the self-similar \emph{ansatz} implies that $\Rm$ and $\LA$ must be functions of $z/R$ (or $\theta$) only. This is not true in full chemical models, except for ambipolar diffusion, which is known to be of order of unity across a wide range of scales \citep[see e.g.][Fig.~8-middle panel]{Thi19}.
In the following, I will therefore mostly focus on models dominated by ambipolar diffusion, and introduce Ohmic diffusion subsequently.

Following \cite{Thi19} models, I prescribe the ambipolar diffusion profile to be
\begin{align}
\label{LAdef}
\LA(z/R)={\LA}_0\exp\Bigg[\frac{z^4}{(\lambda h)^4}\Bigg] 	
\end{align}
where $\lambda$ is a free parameter quantifying the thickness of the non-ideal layer in units of $h$.

In models including Ohmic diffusion, I use the following diffusivity profile
\begin{align*}
\Rm(z/R)={\Rm}_0\exp\Bigg(\frac{z^4}{(\lambda h)^4}\Bigg)\Bigg(\frac{\hat{\rho}(z/h)}{\hat{\rho}(0)}\Bigg)^{-1}.
\end{align*}
This profile, and in particular the dependency on the density ratio, is chosen to be consistent with the ambipolar diffusion profile, assuming the plasma is made of two types of charged species (such as electrons and molecular ions).

\subsection{Numerical method}

In contrast to usual self-similar approaches \citep{Casse00,Jacquemin19} where stationary equation are solved with a shooting methods through the critical points of the flow, here I solve the time-dependent equations (\ref{eq:cont})---(\ref{eq:induct}) using the code PLUTO, a finite-volume, shock-capturing scheme \citep{mignone07}. I use a spherical domain, with the shape of a shell with only one grid point in the radial direction and 2048 points distributed homogeneously in the $\theta$ direction. I choose the grid to be centred on $r=r_0=1$, and the shell extends from $\theta=0.15$ to $\theta=\pi-0.15$. The radial boundary conditions are set enforcing the self-similar relations described in section \ref{sec:ss}. For the boundary in the $\theta$ direction, I use standard outflow boundary conditions. 

The initial condition is a disc in hydrostatic equilibrium threaded by a large scale vertical ($z$) magnetic field whose initial value is set to have $\beta=10^5$ in the disc mid-plane. This initial condition is strongly unstable and the disc very quickly (in less than $10\,\Omega^{-1}$) launches an outflow before reaching a quasi steady-state.

Given that the code is time-dependent and our choice of boundary conditions, the disc mass is not necessarily conserved. Indeed, a fraction of the accreted material can be lost in the wind, leading to a slow decrease of the disc mass. In historical models, this is taken into account by slightly adjusting the power exponent $\gamma_\rho$ to ensure that mass is constant in the disc. This is not possible in our case since the solution is dynamically evolving: one would need to change $\gamma_\rho$ as a function of time. Instead, I therefore choose to renormalise the mass at each time step to keep the total mass inside the domain constant.

For similar reasons, the total magnetic flux threading the disc can evolve on secular timescales in this approach. This is not allowed in usual self-similar approaches, which usually assume that the total toroidal electromotive field is null. Here, I choose not to enforce such a constraint. This allows me to measure the transport of magnetic flux, but it also implies that the magnetic flux evolves with time. Hence, as for the density, I multiply the each component of the field by a fixed factor at each time step, adjusted so that the magnetic flux threading the disc is equal to the desired value. 

\subsection{Accretion theory and diagnostics\label{sec:diags}}

Several diagnostics can be derived from self-similar solution. The most useful diagnostics are related to accretion theory, i.e. how the disc surface density and mass accretion rate evolve with space and time. Let me therefore develop the mass, angular momentum and magnetic flux conservation equations as:
\begin{align}
\label{eq:acc_mass}
\frac{\partial \Sigma}{\partial t}+\frac{1}{R}\frac{\partial}{\partial R}\frac{\Macc}{2\pi}&=	\Big[\overline{\rho v_z}\Big]_{-z_0}^{z_0},\\
\label{eq:acc_ang}
\frac{\Macc v_\mathrm{K}}{4\pi}&=\frac{\partial }{\partial R}\left(R^2\int_{-z_0}^{z_0}\mathrm{d}z\,\overline{T_{R\varphi}}\right)+R^2\Big[\overline{T_{z\varphi}}\Big]_{-z_0}^{z_0},\\
\label{eq:acc_flux}
\frac{\partial \overline{B_{z0}}}{\partial t}&=-\frac{1}{R}\frac{\partial }{\partial R} R \overline{\mathcal{E}_{\varphi 0}},
\end{align}
where the overline denotes an azimuthal and ensemble average (in the case of self-similar stationary solution, this average is not strictly needed), $\Sigma=\int_{-z_0}^{z_0}\mathrm{d}z\,\overline{\rho}$ is the disc surface density, $z_0$ is the disc surface, $\Macc\equiv-2\pi R\int_{-z_0}^{z_0}\mathrm{d}z\,\overline{\rho v_R}$ is the mass accretion rate, $T_{x\phi}=\rho v_x(v_\varphi-v_\mathrm{K}) -B_xB_\varphi/4\pi$ is the stress $\varphi$ component while $B_{z0}=B_z(z=0)$ and $\mathcal{E}_{\varphi 0}$ are the vertical magnetic field strength and azimuthal EMF in the disc plane, respectively. It should be noted that in principle, accretion theory only require the first two relations. However, it is now well established that in a wind-driven disc, stresses and mass loss rates are also function of the mean poloidal field strength, hence, in this context, accretion theory has to be supplemented by (\ref{eq:acc_flux}).

Although equations (\ref{eq:acc_mass})---(\ref{eq:acc_flux}) fully describes the secular evolution of any disc at hand, it should be realised that the right hand-side terms are a priori unknown. This closure problem is well known in the disc community, and is usually solved by using the $\alpha$ disc paradigm in cases where only the $T_{r\varphi}$ term is present. In the case of wind-driven disc, 4 terms are actually present. Following the $\alpha$ disc idea, I therefore define 4 parameters:
\begin{align}
	\zeta_\pm&\equiv \pm\frac{\overline{\rho v_z}(\pm z_0)}{\Sigma \OK},\\
	\alpha&\equiv\frac{\int_{-z_0}^{z_0}\mathrm{d}z\,\overline{T_{R\varphi}}}{\int_{-z_0}^{z_0}\mathrm{d}z\overline{P}}=\frac{\int_{z_0}^{z_0}\mathrm{d}z\,\overline{T_{R\varphi}}}{\Sigma \OK^2h^2},\\
	\upsilon_{\pm}&\equiv\pm\frac{\overline{T_{z\varphi}}(\pm z_0)}{\Sigma\OK^2 h},\\
	v_B &\equiv \frac{\overline{\mathcal{E}_{\varphi 0}}}{\Omega_K h \overline{B_{z0}}},
\end{align}
where I have assumed that the disc was isothermal to define $\alpha$. With these definitions, a positive $v_B$ implies that the field is transported \emph{outwards}.
Note that the mass loss rate $\zeta$ is comparable to the definition of \cite{Scepi18}, up to a factor of order unity. In numerical applications, I use $z_0=6h$.

It is also customary to measure the magnetic field strength as a function of the plasma $\beta$ parameter. In this manuscript, I will define the plasma $\beta$ on the strength of the mean vertical field threading the disc mid-plane, i.e.
\begin{align}
\beta\equiv \frac{8\pi \overline{\rho}(z=0) c_s^2}{\overline{B_{z0}}^2}.
\end{align}

The transport coefficients $\zeta$, $\alpha$, $\upsilon$ and $v_B$ being dimensionless coefficients, they are expected to depend only on dimensionless numbers. We therefore expect them to depend on the magnetic field strength $\beta$, but also on the diffusivity of the disc $\Rm$ and $\LA$. In principle, once all of these dependencies are known, accretion theory is said to be \emph{complete} and one can compute the evolution of a wind-driven protoplanetary disc in a way similar to the historical $\alpha$ disc. Therefore, determining these dependencies is the main objective of this article.

\section{Fiducial simulation}
\subsection{Time evolution}
\begin{figure}
   \centering
   \includegraphics[width=1.1\linewidth]{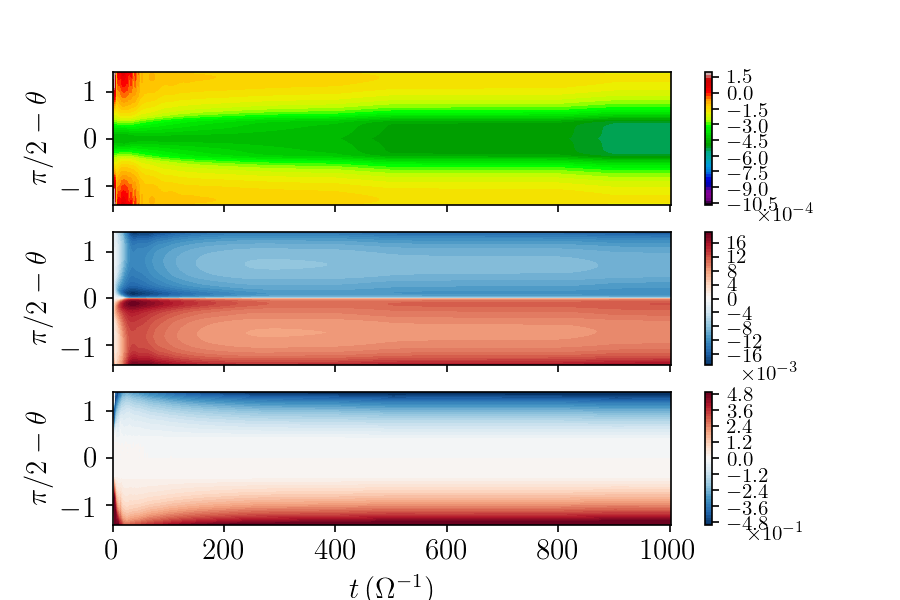}
      \caption{Space-time diagram showing the evolution for the first $1000\,\Omega^{-1}$ of $B_\theta$ (top), $B_\varphi$ (middle) and $v_\theta$ (bottom) for the fiducial run. The system quickly reaches a steady-state. In order to continue the solution as a function of the field strength, the field is slowly increased every $300\,\Omega^{-1}$.
              }
         \label{fig:spacetime}
 \end{figure}

\begin{figure*}
   \centering
   \includegraphics[width=0.32\linewidth]{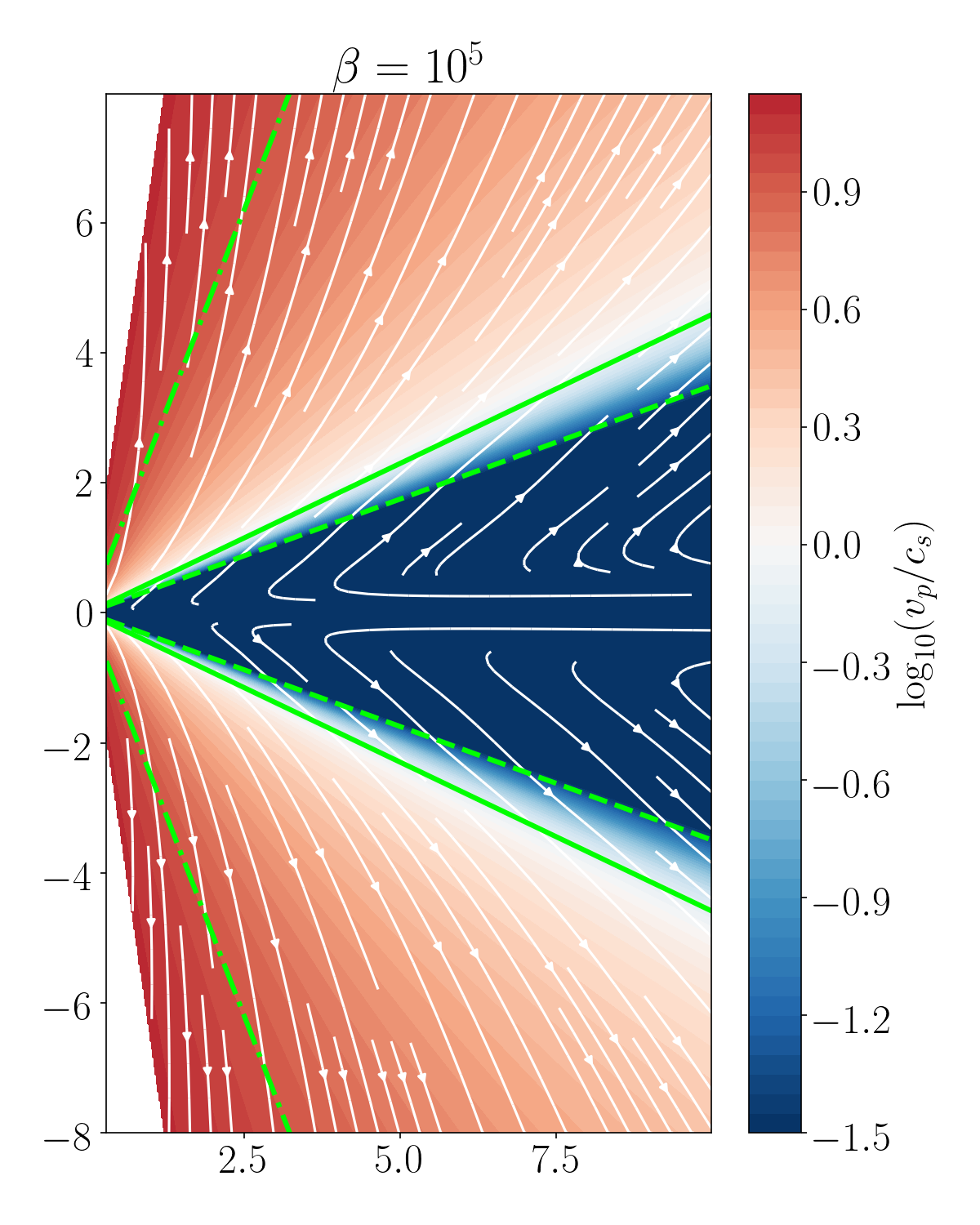}
   \includegraphics[width=0.32\linewidth]{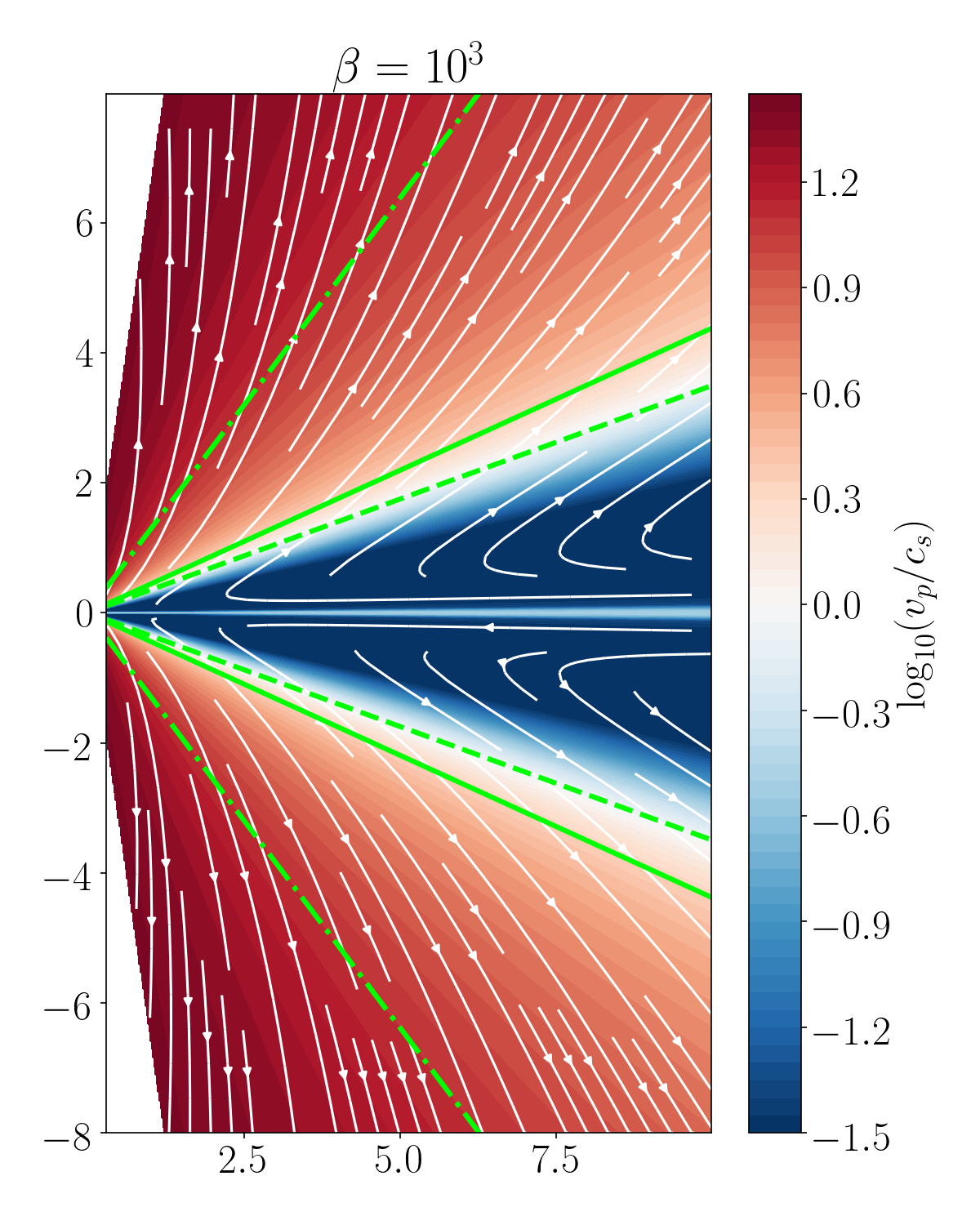}
   \includegraphics[width=0.32\linewidth]{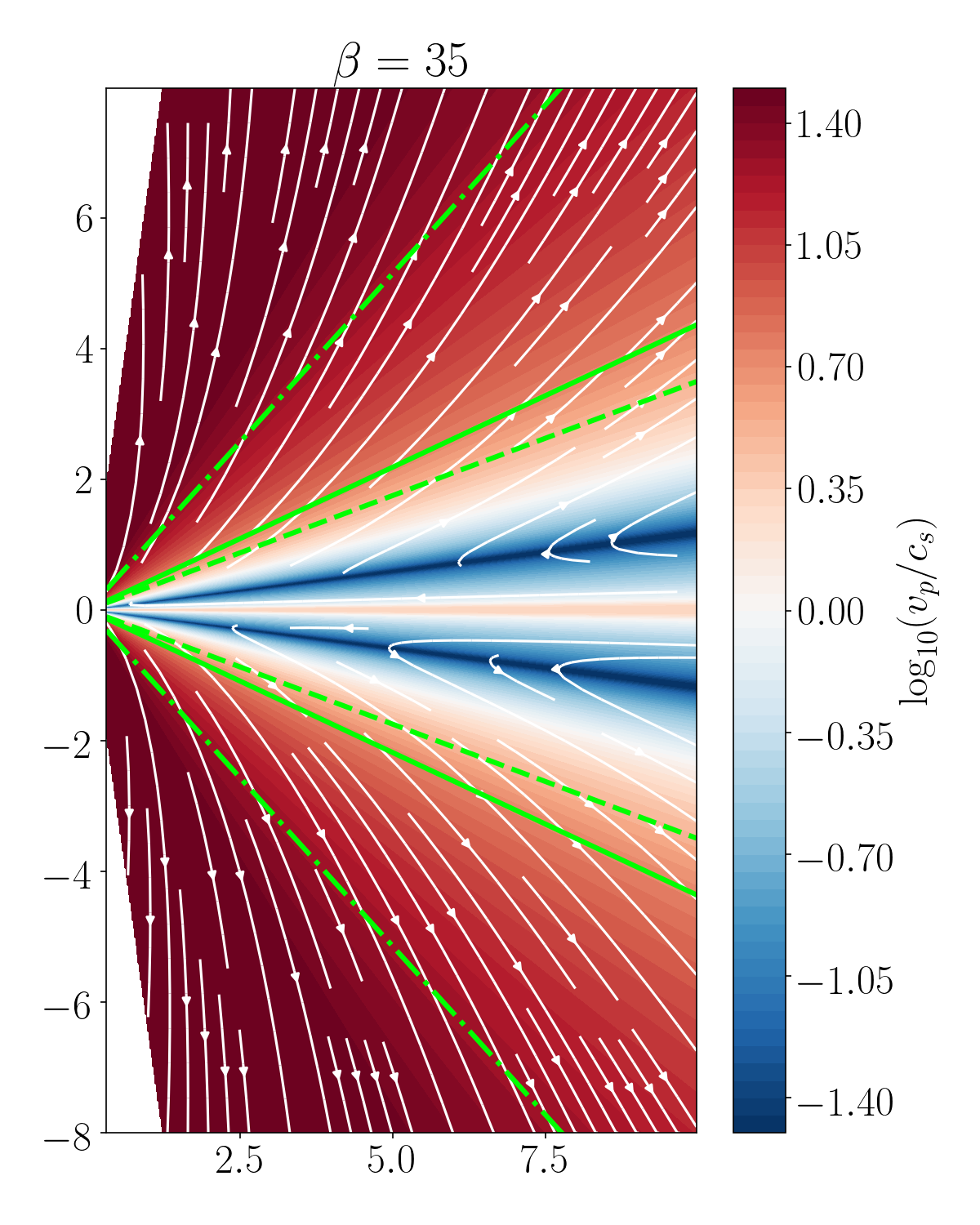}
   
   \includegraphics[width=0.32\linewidth]{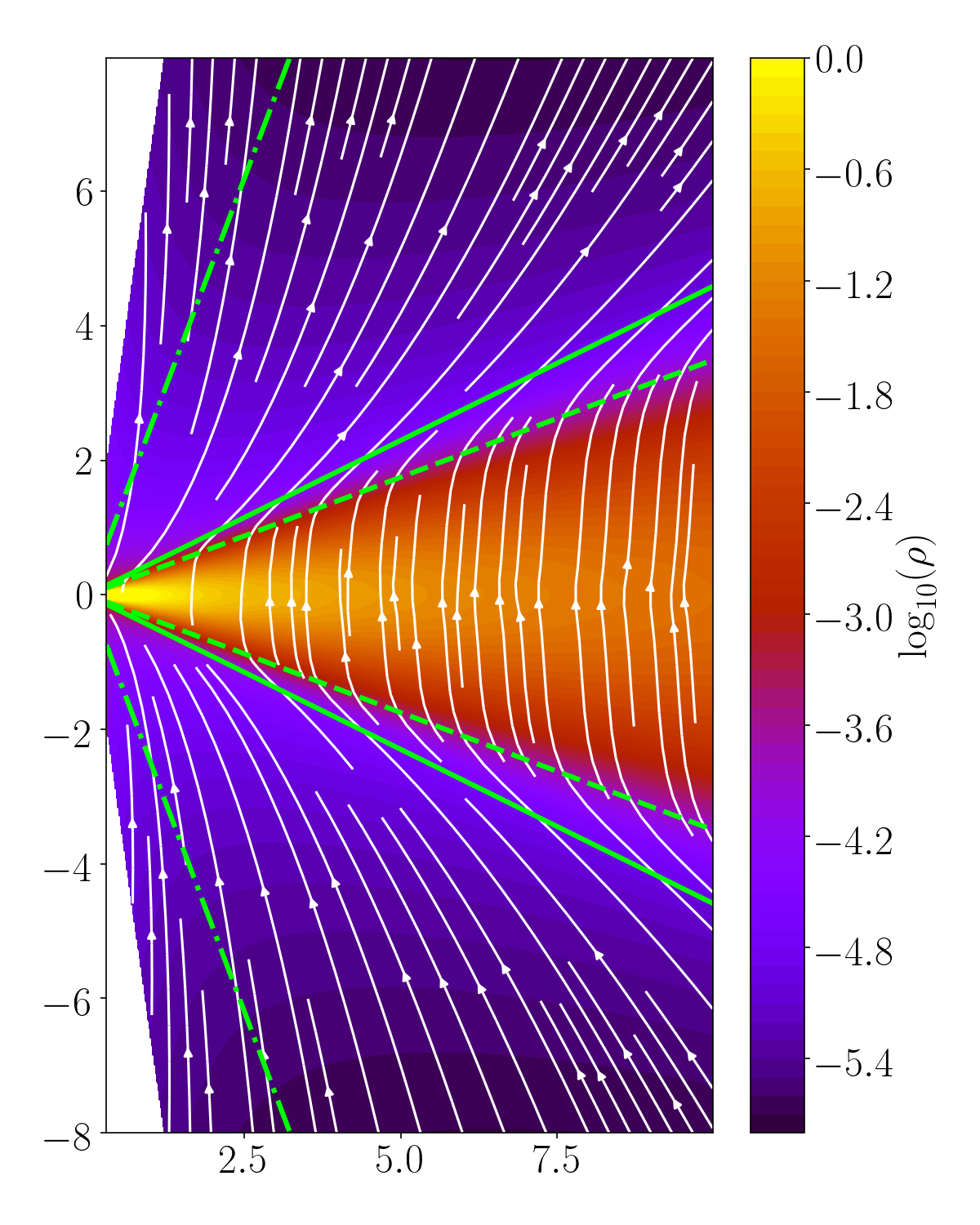}
   \includegraphics[width=0.32\linewidth]{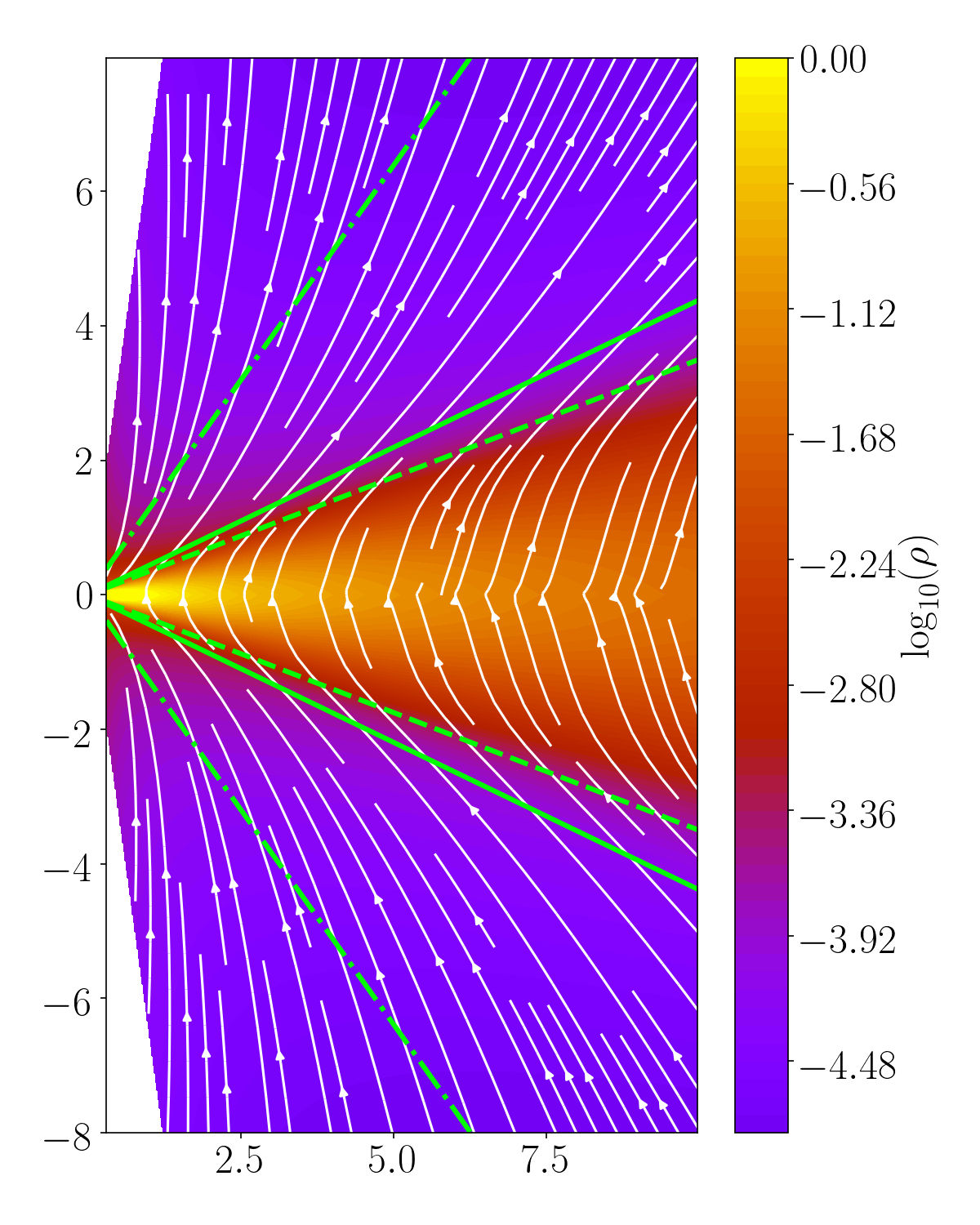}
   \includegraphics[width=0.32\linewidth]{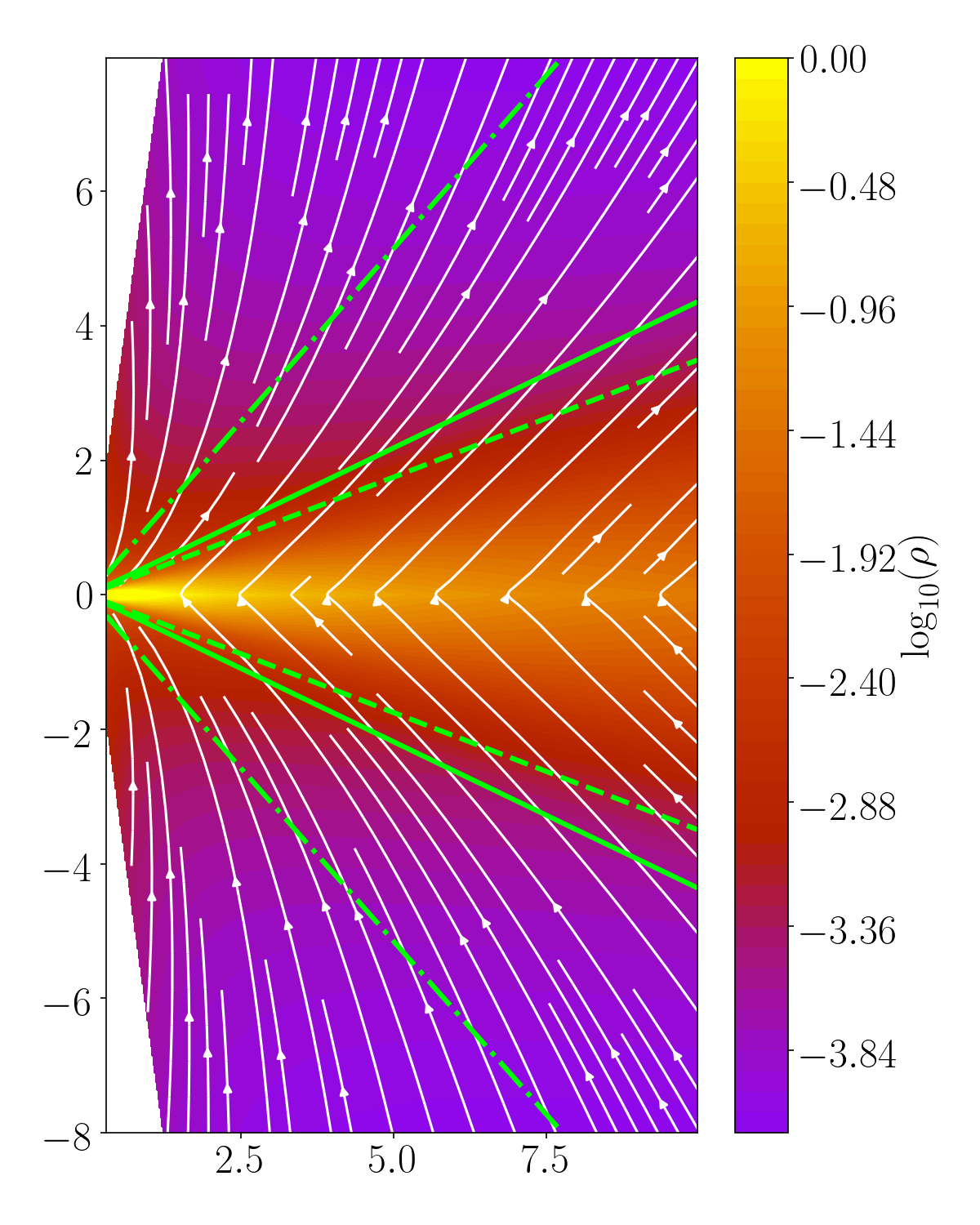}
      \caption{Flow topology in the fiducial run (ambipolar diffusion only $\LA=1$). Top row: poloidal streamlines (white) and log of the sonic Mach number.  Bottom row: poloidal field and $\log$ of density, normalised so that the mid-plane density at $R=1$ is unity. Note that the colour scales are identical between the columns. From left to right the disc magnetisation increases: $\beta=10^5\,;\,10^3\,;\,35$.  The green lines denotes critical lines of the flow: Alfv\'enic (plain) and fast magnetosonic (dot-dashed). The green dashed line represents the disc "surface" where the flow becomes ideal, arbitrarily located at $z=3.5h$ for all of the solutions.
              }
         \label{fig:flowFiducial}
   \end{figure*}

In the following, I will define the fiducial run as a simulation with ambipolar diffusion only, $\LA=1$ and $\lambda=3$ so that the non-ideal MHD zone extends up to $3h$ above the disc mid-plane. The disc aspect ratio is fixed to $\varepsilon=0.1$ and the disc and atmosphere are assumed to be locally isothermal, i.e. $c_s(R,Z)=c_s(R,z=0)$. As for all the simulations, the fiducial run is started with a mid-plane $\beta=10^5$. Note that I have also ran this simulation at half resolution (i.e. 1024 points in $\theta$) to check the convergence of the solution. The two solutions differ by less than $1\%$, so I am confident that the full resolution simulations presented here are perfectly resolved.

While most of the solutions presented in this article are quasi-steady state, let me briefly discuss the space-time evolution of the simulations used to compute the solutions. Such a space-time evolution is shown in Fig.~\ref{fig:spacetime}, for the fiducial run. As it can be seen, the system rapidly relaxes into a steady-state. During the first $100\,\Omega^{-1}$, I apply a linear damping to the equations of motion to damp epicyclic oscillations more rapidly than the system would naturally do. These oscillations are due to the sudden launching of the wind, and are a spurious result of my wind-free initial conditions (a similar procedure was used in shearing box by \citealt{lesur13}). 

After the first $100\,\Omega^{-1}$, the system reaches a quasi-steady state, from which I can compute the information needed: angular momentum transport, mass flux, etc. In order to continue the solution as a function of $\beta$, I slowly increase the total field strength between $300\,\Omega^{-1}$ and $400\,\Omega^{-1}$, and let the system reach its new equilibrium for the next $200\,\Omega^{-1}$. I then repeat this procedure over and over, until the code stops because of singular points in the solution. In doing so, I reached $\beta=35$ in the fiducial run, by steps of $10\%$ decrease in $\beta$. In the fiducial setup, I have also continued the solutions from $\beta=10^5$ up to $\beta=10^{8}$ by steps of $10\%$ \emph{increase} to study the domain of validity of the solutions obtained by this procedure. Hence, the domain explored for the fiducial setup is wider than for the other solutions (section \ref{sec:parameters}).

In the following, I average the quasi-steady flow obtained for the last $100\Omega^{-1}$ of "relaxation" for each magnetisation and use the result to compute the flow properties.

\subsection{Flow topology}

The 2D topology of the flow can be deduced from the 1D "shell" solutions of the simulation by reconstructing the 2D fields using the self-similar scalings (\ref{eq:ssDefinition}). Using this procedure, I compute the flow topology for the two extreme cases ($\beta=10^5$ and $\beta=35$) and an intermediate case ($\beta=10^3$) shown in Fig.~\ref{fig:flowFiducial}. In this fiducial simulation, the flow is symmetric with respect to the disc mid-plane, as seen here. The strongly magnetised solution produces a super-sonic accretion flow, while the weaker magnetised case tend to produce subsonic accretion ($v_p/c_s<0.1$). In the weak field case, the field lines are almost straight, indicating the the decoupling between the field and the flow is efficient, thanks to ambipolar diffusion. As the field gets stronger case however, the field lines are clearly pinched around the mid-plane. This indicates a strong field dragging by the accretion flow.

The wind is in all cases super fast-magnetosonic. The fast surface is closer to the disc surface when the magnetisation gets larger, while the Alfv\'en surface tends to stay at the same altitude (around $4.5h$). Generally speaking, stronger magnetisation leads to faster and more massive winds (for an identical disc). Interestingly, while the shape of the stream and field lines differ significantly in the disc, their aspect is very similar in the wind region starting above the disc surface. Note that in all of these cases, the flow show signs of collimation towards the axis. 

\subsection{Transport properties}

\begin{figure*}[h!]
   \centering
   \includegraphics[width=0.95\linewidth]{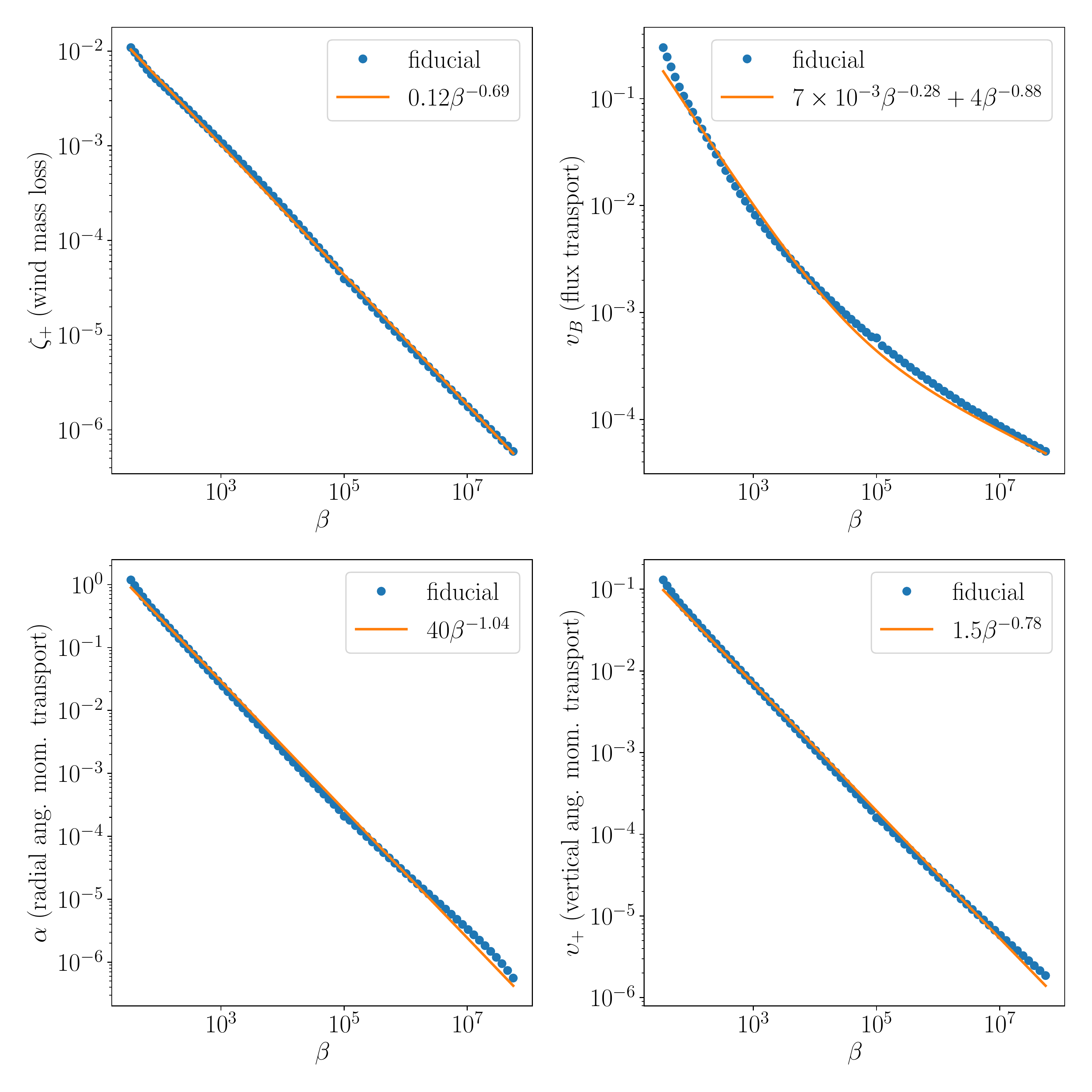}

      \caption{Transport coefficient as a function of the disc magnetisation for the fiducial run (ambipolar diffusion only ${\LA}_0=1$). Each blue dot is a steady state solution obtained by continuation in the simulation. Because the flow is symmetric with respect to the mid-plane, only values computed from the disc top surface ($\zeta_+$ and $\upsilon_+$) are shown. The best fits using power laws are shown in plain orange line.}
         \label{fig:statFiducial}
   \end{figure*}
   
The transport properties defined in section \ref{sec:diags} can be evaluated as a function of the magnetisation for the fiducial run. This gives the dependencies shown in Fig.~\ref{fig:statFiducial}. The resulting transport coefficients are remarkably close to power laws of $\beta$ (except for $v_B$ which shows a shallower dependence on $\beta$ than a power law). Several important trends can be already deduced from this result.

First, it should be pointed out that since the solutions are all steady-state and laminar, the $\alpha$ measured here is a \emph{pure laminar stress}. It is not due to any form of turbulence, as it is just the result of the large scale wind which transports angular momentum \emph{both} radially and vertically. Still, it can be compared to what is found in turbulent discs. The dependence of $\alpha$ on $\beta^{-1}$ is steeper than the one usually found in ideal-MHD shearing box simulations, which usually give $\alpha\propto \beta^{-1/2}$ \citep[e.g.][]{salvesen16, Scepi18}. Note however that the ideal and non-ideal values for $\alpha$ match for $\beta\sim 10$. Hence, my values of $\alpha$ are generally smaller than their ideal MHD counterpart by a ratio $\sim \beta^{1/2}/3$. This is most likely the signature of ambipolar diffusion in the disc. 

In general, the values for $\alpha$, $\upsilon$ and $\zeta$ are compatible with the values found in the shearing box literature for $\beta\sim 10^4-10^5$ and equivalent diffusivity regimes dominated by $\LA=1$ \citep[e.g.][]{Bai13,lesur14}, even though the dependency on $\beta$ are slightly steeper than \cite{Bai13}.

\begin{figure}
 	\centering
   \includegraphics[width=0.95\linewidth]{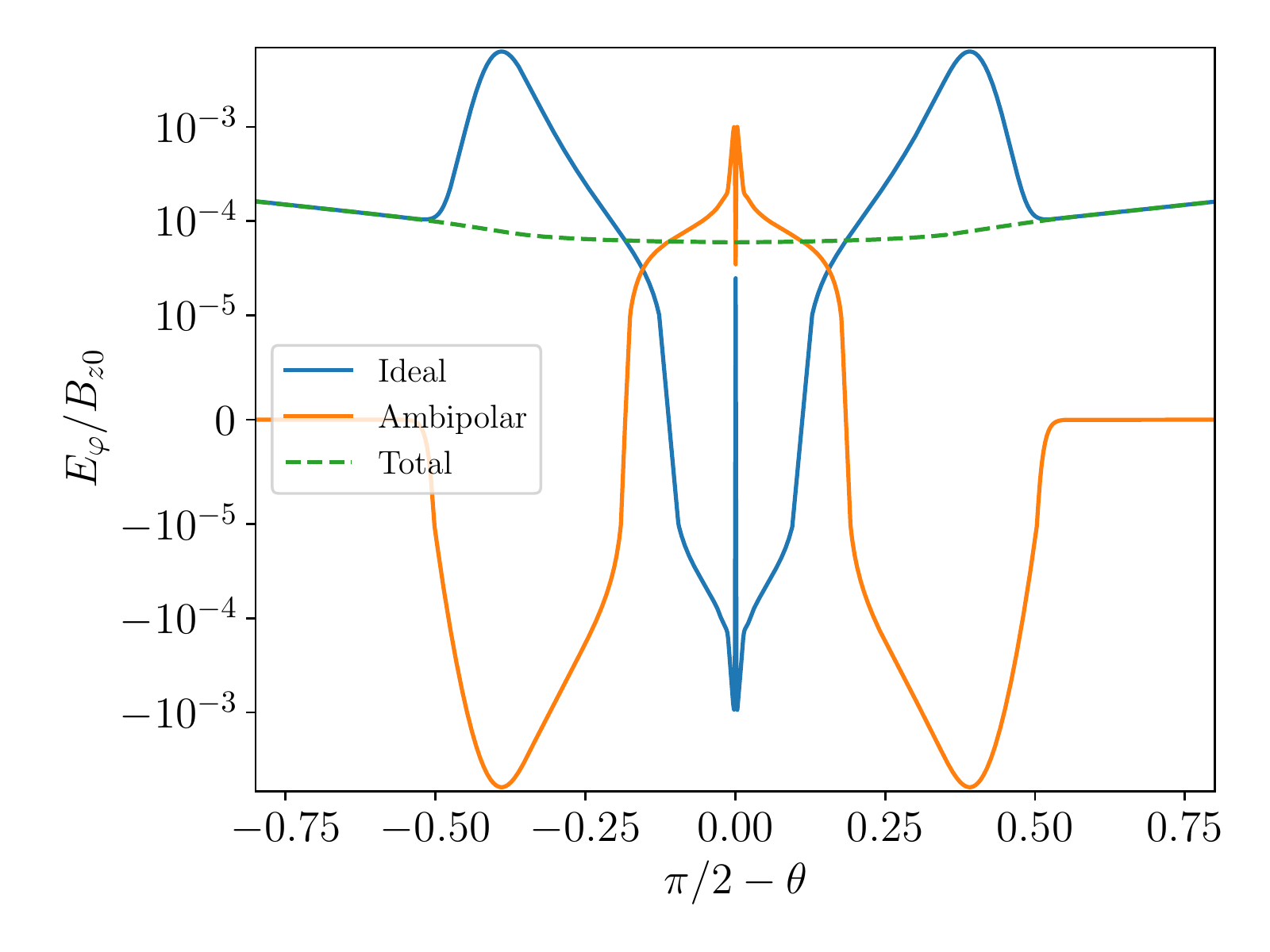}

      \caption{Zoom on the toroidal electromotive field of the disc, splitting the ideal MHD and ambipolar diffusion contributions for $\beta=10^5$ in the fiducial case. Note that the two electromotive fields almost exactly balance. The total EMF at $\theta=\pi/2$ is $\mathcal{E}_\varphi/B_{z0}\simeq 6\times 10^{-5}$ and is approximately constant throughout the disc. This gives $v_B=6\times 10^{-4}$.}
         \label{fig:Ephi}
 \end{figure}

Interestingly, $v_B>0$ for all the cases explored in the fiducial run. This indicates that despite the strong accretion in the disc mid-plane, the field is always \emph{diffusing outwards}. This is qualitatively consistent with \cite{BaiStone17}, but the diffusion \emph{speeds} are reduced by an order of magnitude in my case. For $\beta=10^4$, I get $v_B=1.7\times 10^{-3}$ while \cite{BaiStone17} get\footnote{Note that I measure $v_B$ in units of $c_s$ while \cite{BaiStone17} quote $v_B$ in units of $v_K$, hence a factor of $\varepsilon$ should be added when comparing the two results.} $v_B=4\times 10^{-2}$. A careful examination of the ideal and non-ideal contributions to $\mathcal{E}_\varphi$ (Fig.~\ref{fig:Ephi}) shows that advection almost completely balance diffusion, diffusion winning only by a 1\% excess. It is also worth noting that while the ideal and ambipolar EMFs vary strongly in the disc, their sum is approximately constant with $\theta$, indicating that poloidal field lines are ``moving'' radially without much vertical deformation in the disc.

Using eq.~\ref{eq:acc_ang}, the safe-similar scalings (eq.~\ref{eq:ssDefinition}) and the definitions of section \ref{sec:diags}, it is straightforward to show that the mass accretion rate is directly related to $\alpha$ and $\upsilon_{\pm}$
\begin{align*}
\frac{\Macc}{2\pi \Sigma R^2\Omega_K}\equiv\dot{m}=\varepsilon\Big(\underbrace{\alpha\varepsilon}_{\dot{m}_R} +\underbrace{2(\upsilon_++\upsilon_-)}_{\dot{m}_z}\Big)	
\end{align*}
where I have split $\Macc$ in a contribution from the radial ($\dot{m}_R$) and vertical ($\dot{m}_z$) stresses. Even though these are usually referred to as "turbulent" and "wind" contributions (even in cases where the flow is laminar), I emphasise that there is no turbulent stress. Hence both contributions are intrinsically due to the wind. Using the power law scalings in Fig.~\ref{fig:statFiducial}, I find that
\begin{align*}
\frac{\dot{m}_z}{\dot{m}_R}=1.5	\beta^{0.26}.
\end{align*}
Hence whenever $\beta\gg 1$, the vertical stress is largely dominant and the radial transport of angular momentum can be neglected altogether, which is consistent with previous 3D numerical simulations \citep{Bethune17}.

Neglecting $\dot{m}_R$, I can derive the mass accretion rate in a realistic disc using only the scaling for $\upsilon_\pm$, which gives
\begin{align}
\nonumber \Macc = 6.6\times 10^{-8}\left(\frac{\Sigma}{10\,\mathrm{g.cm}^{-2}}\right)&	\left(\frac{R}{10\,\mathrm{A.U.}}\right)^{0.5}\times \\
\label{eq:mdotfit}\quad &\left(\frac{M}{M_\odot}\right)^{0.5}\left(\frac{\beta}{10^4}\right)^{-0.78}\,M_\odot/\mathrm{yr},
\end{align}
or substituting $\beta$ with the field strength and the surface density
\begin{align}
\nonumber \Macc = 1.6\times 10^{-8}&\left(\frac{\Sigma}{10\,\mathrm{g.cm}^{-2}}\right)^{0.22}	\left(\frac{R}{10\,\mathrm{A.U.}}\right)^{2.08}\left(\frac{M}{M_\odot}\right)^{-0.28} \\
\label{eq:mdotB}\quad &\qquad\times \left(\frac{\varepsilon}{0.1}\right)^{-0.78}\left(\frac{B_z}{1\,\mathrm{mG}}\right)^{1.56}\,M_\odot/\mathrm{yr}.
\end{align}

It is worth noting that the mass accretion rate does not depend strongly on the disc surface density, in contrast to the usual viscous disc model. Instead, I find that it is the field strength that is the dominant control parameter in these solutions. More quantitatively, I find accretion rates compatible with expected ones from observations ($\sim 10^{-8}$---$10^{-7}\,M_\odot/\mathrm{yr}$) when $\beta\in[10^3,10^5]$ or $B_z\sim$ a few mG. A word of caution though: accretion rates are typically inferred from the material which falls onto the central star, while here the mass accretion rate is computed in the disc at 10 AU. If the wind extract a lot material between the star and 10 AU, the accretion rates have no reason to match. 

A common way to quantify what fraction of the accreted mass is ejected is through the ejection efficiency parameter $\xi$ defined as
\begin{align*}
\xi&\equiv\frac{2\pi R^2\Big[\overline{\rho v_z}\Big]_{-z_0}^{z_0} }{\Macc}= \frac{\mathrm{d}\log \Macc }{\mathrm{d}\log R}	\quad\textrm{in steady state}
\end{align*}
where the last equality results from the continuity equation assuming $\partial_t\Sigma =0$. It is easy to show that the ejection efficiency is directly connected to the transport coefficients:
\begin{align*}
\xi&=\frac{\zeta_++\zeta_-}{2\varepsilon (\upsilon_++\upsilon_-)}\\
&=	0.4\,\beta^{0.09}
\end{align*}
where the first equality assumes $\dot{m}_R$ is negligible while the second uses the scaling found in the fiducial run (Fig.~\ref{fig:statFiducial}). This implies that $\xi=O(1)$ for $\beta\sim 10^4$. Therefore, the accretion rate is approximately a linearly increasing function of radius in steady-state in these models!

Finally, while most of the scalings presented here use $\beta$ and therefore the vertical field strength as the main control parameter, these solutions are actually dominated by the azimuthal field component $B_\varphi$. Using the definition of $\upsilon$ it easy to show that
\begin{align}
\nonumber\left|\frac{B_\varphi}{B_z}\right|_{z_0}&=\sqrt{\frac{\pi}{2}}\beta \upsilon\\
\label{eq:magshear} &\simeq 	1.9\beta^{0.22},
\end{align}
where the last equality comes from the scaling of the fiducial run. Hence, for $\beta=10^4$, $|B_\varphi|$ is  more than 14 times larger than $B_z$ at the disc surface, indicating that the field is strongly wrapped at that location. Note finally that since the $\upsilon_+=-\upsilon_-$, the sign of $B_\phi$ is reversed on the top and bottom side of the disc.
\subsection{Magnetic torque and currents\label{sec:current}}

 \begin{figure}
 	\centering
   \includegraphics[width=0.95\linewidth]{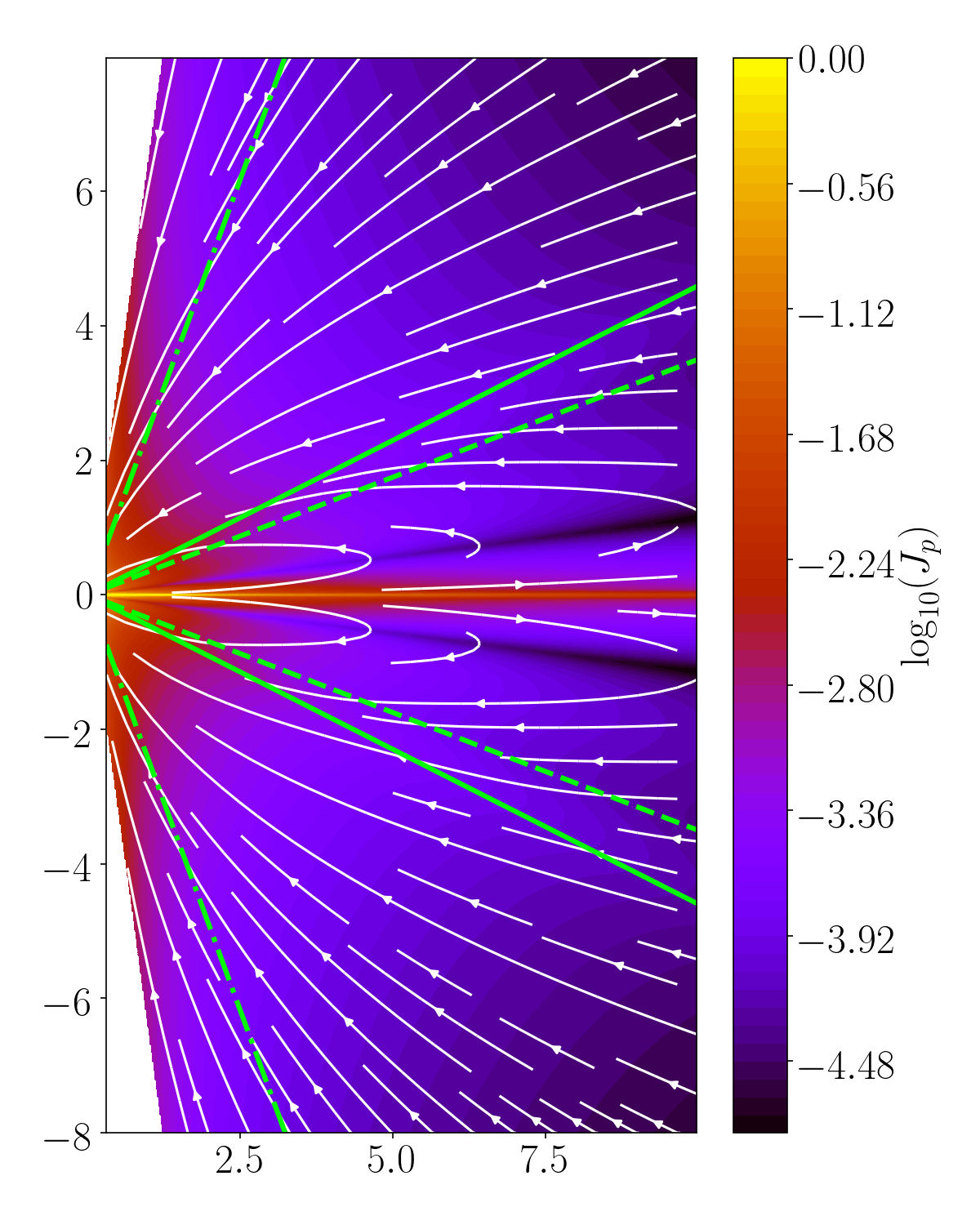}

      \caption{Poloidal current in the fiducial simulation at $\beta=10^5$. The current intensity is represented in colour. Note the presence of a strong outwards current in the disc mid-plane. The green lines represents the same characteristics as Fig.~\ref{fig:flowFiducial}. }
         \label{fig:current}
 \end{figure}

Although it is possible to interpret accretion in terms of angular momentum budget (the disc angular momentum is transported into the wind), this gives us little information about \emph{how} the disc is actually accreting. This question can be addressed by looking at the forces acting on the disc. In particular, because the outflow is magnetised, the Lorentz force $\bm{F}_L=\bm{J\times B}/c$ is key. If $F_{L,\varphi}<0$, the Lorentz force breaks down the disc rotation by creating a negative torque, driving accretion. One easily sees that the magnetic torque is closely related to the poloidal current
\begin{align*}
	\bm{F}_{L,\,\varphi}=\frac{\bm{J}_p\times \bm{B}_p}{c}.
\end{align*}
Because the poloidal field is more homogeneously distributed in the disc (Fig.~\ref{fig:flowFiducial}), the magnetic torque is strongest in the regions of strong $\bm{J}_p$. Hence the poloidal current is a direct indicator of the torque exerted by the wind on the disc. The distribution of poloidal currents in the fiducial $\beta=10^5$ case is shown in Fig.~\ref{fig:current}.  I find a strongly focussed current sheet in the disc mid-plane, with currents directed outwards. It is this current which induces a strong negative torque, and which, in turn, is responsible for accretion in the disc mid-plane. This current then closes back in the disc surface and in the wind.

\begin{figure}
 	\centering
   \includegraphics[width=0.95\linewidth]{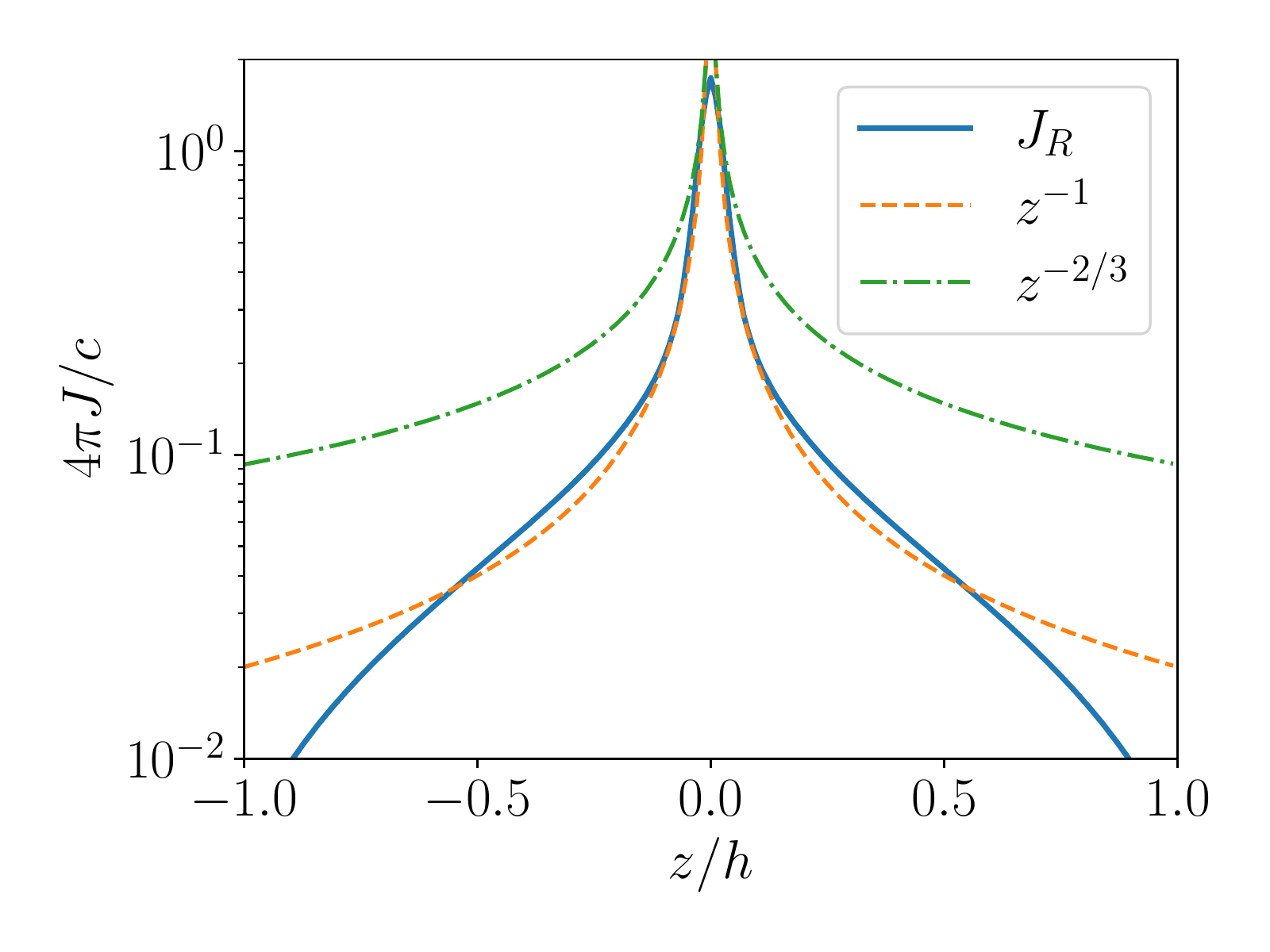}

      \caption{Zoom on the radial current in the fiducial simulation at $\beta=10^5$. The current sheet is strongly focused, with a scaling compatible with $J_r\propto z^{-1}$ (see text). }
         \label{fig:currentZoom}
 \end{figure}

It might be surprising at first sight to find such a narrow current sheet in the system given that the disc is strongly diffusive ($\LA=1$). It is however well known that ambipolar diffusion is a self-focusing diffusion process. It's been argued that in "free" reconnection sites, the current scales like $z^{-2/3}$, where $z$ is the distance from the null point \citep{Zweibel97}. In the particular case presented here, I find that the current is even more focused with $\bm{J_r}\propto z^{-1}$ close to the mid-plane. Note that it is not so surprising since in the present case, we're dealing with a reconnection site "forced" by the outflow launched from the surface, hence the scaling need not be identical to that of \cite{Zweibel97}.

\subsection{Wind properties}
Winds in protoplanetary discs are often refereed to as magneto-thermal winds. This is because most (if not all) of the global simulations published up to now have been using a "hot atmosphere". These hot atmospheres are either achieved by a prescribed heating function  \citep{Bethune17}, or with a more complete treatment of the thermodynamics including various sources of heating and cooling \citep{Wang19,Gressel20}. In any case, the energetics of these winds is partially dominated by thermal heating, hence their name. 

In the fiducial solution presented here, the equation of state is locally isothermal. Hence the temperature of the atmosphere is that of the disc at the same cylindrical radius $R$. For that reason, my fiducial wind is $\emph{not}$ magneto-thermal, as we shall see.

\begin{figure}
 	\centering
   \includegraphics[width=0.95\linewidth]{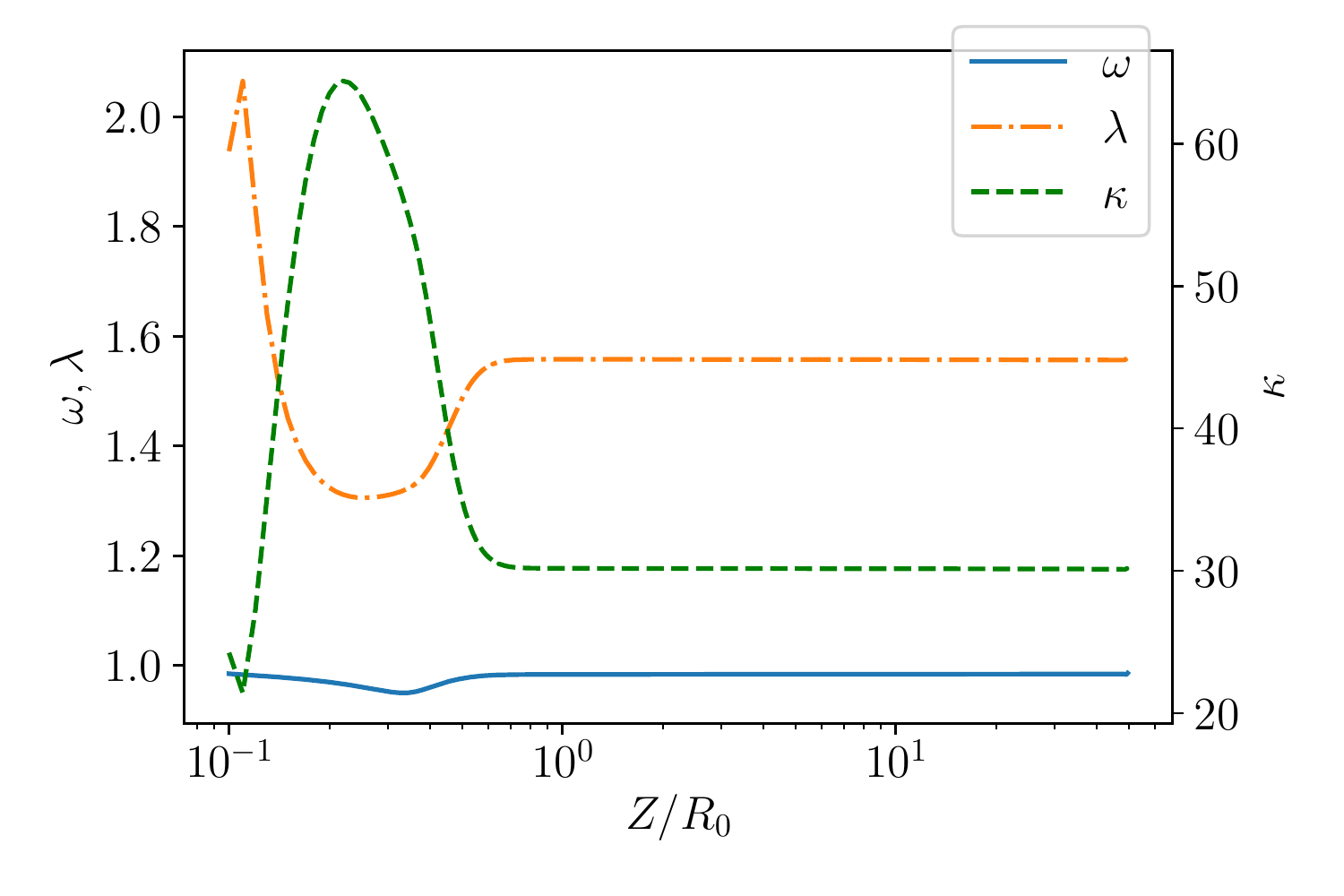}

      \caption{MHD invariants plotted along one given field line for the fiducial run at $\beta=10^5$. As expected, the invariants are approximately constant once in the ideal-MHD region above the disc. The asymptotic values are $\kappa=30.2,\,\omega=0.99$ and $\lambda=1.56$.}
         \label{fig:invariants}
 \end{figure}

\begin{figure}[h!]
 	\centering
   \includegraphics[width=0.95\linewidth]{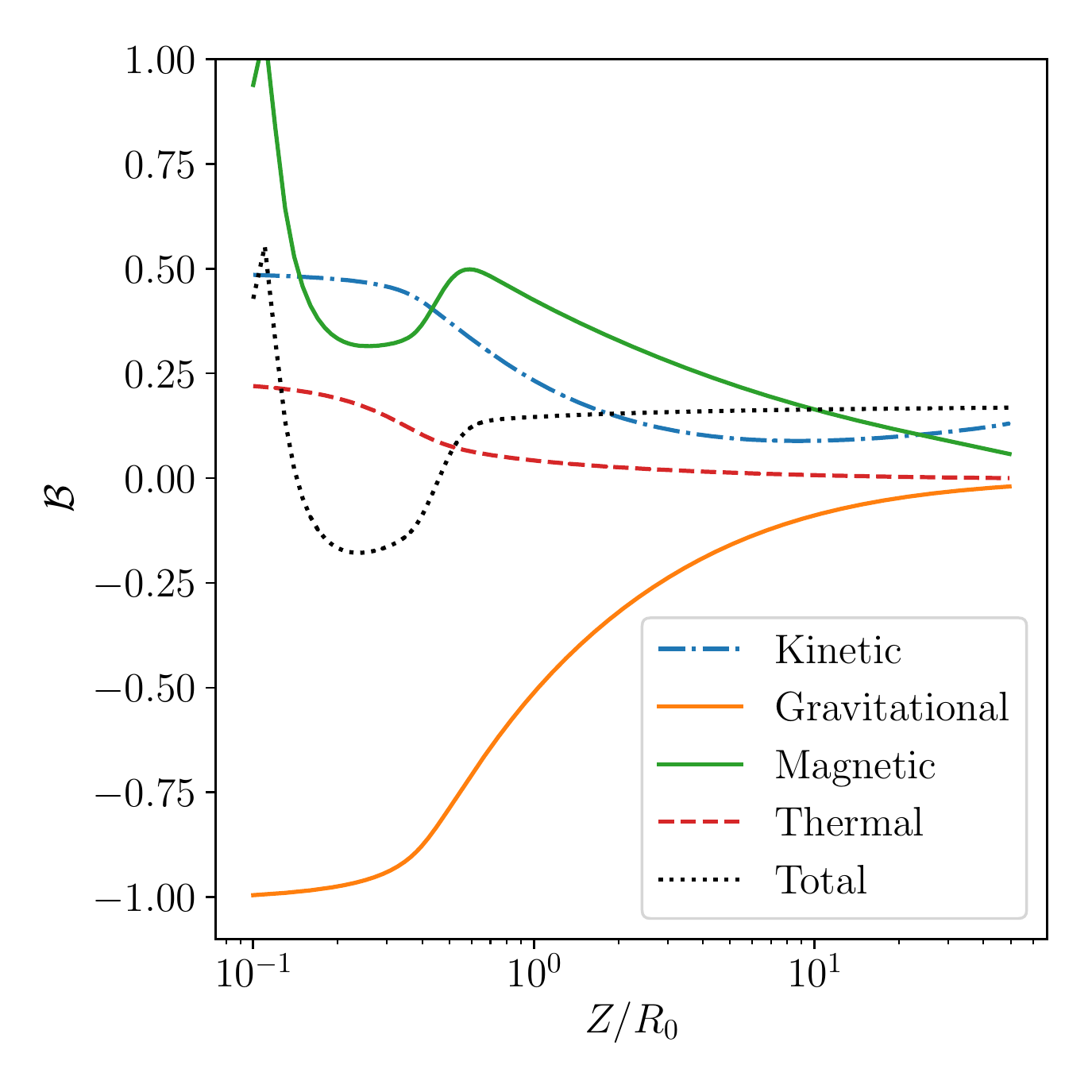}

      \caption{Bernoulli invariant plotted along one given field line for the fiducial run at $\beta=10^5$. The Bernoulli invariant becomes constant once in the ideal MHD region, as expected. The thermal contribution to the energetics is negligible compared to the magnetic contribution, hence the wind is classified as a "cold" MHD wind. }
         \label{fig:bernoulli}
 \end{figure}

Since the wind is a steady-state ideal MHD flow, several conserved quantities can be derived from the equations of motion. Let me therefore consider a streamline passing through the mid-plane at $R_0$. I define $\Omega_0$ as the Keplerian angular velocity at $R_0$: $\Omega_0\equiv \Omega_\mathrm{K}(R_0)$. I then follow \cite{blandford82} defining:
 
\begin{itemize}
\item The mass loading parameter 
\begin{align*}
\kappa=\frac{4\pi \rho u_p\Omega_0R_0}{B_pB_0},	
\end{align*}
where $v_p$ and $B_p$ are the poloidal velocity and field strength, quantifies the amount of mass loaded onto the field lines.
\item The rotation parameter 
\begin{align*}
\omega\equiv \frac{\Omega}{\Omega_0}-\frac{\kappa B_0 B_\varphi}{4\pi \rho R R_0 \Omega_0^2}	
\end{align*}
which measures the rotation speed of magnetic surfaces
\item The magnetic lever arm
\begin{align*}
\lambda\equiv\frac{\Omega R^2}{\Omega_0R_0^2}-\frac{R B_\varphi}{R_0\kappa B_0}	
\end{align*}
which measures the amoung of angular momentum extracted by the wind
\item The Bernoulli invariant
\begin{align*}
\mathcal{B}=\underbrace{\frac{v^2}{2\Omega_O^2R_0^2}}_\mathrm{Kinetic}\underbrace{-\frac{R_0}{\sqrt{R^2+z^2}}}_\mathrm{Gravitational}\underbrace{-\frac{\omega R B_\varphi}{\kappa R_0 B_0}}_\mathrm{Magnetic}\underbrace{+\frac{w}{\Omega_0^2 R_0^2}}_\mathrm{Thermal}
\end{align*}
measures the energy content of the flow. In this definition, I have included the thermal contribution to the flow energetics as the work done by thermal pressure forces along the field line
\begin{align*}
w\equiv \int_s^\infty -\mathrm{d}\bm{\ell}\cdot \frac{\bm{\nabla}P}{\rho}. 	
\end{align*}
If the flow is adiabatic, this is simply the enthalpy. However, the locally isothermal approximation does not lead to an adiabatic flow, hence I have to use the integral form.
\end{itemize}
I have computed the mass loading, rotation and lever arm parameters along a magnetic field line for the fiducial simulation at $\beta=10^5$(Fig.~\ref{fig:invariants}). As expected, these parameters are constant once above the non-ideal region of the disc. Surprisingly, the lever arm is very small, reaching only $\lambda=1.56$. As is well known, cold MHD winds require $\lambda>3/2$, so we are just above this limit value. A close inspection at the Bernoulli invariant (Fig.~\ref{fig:bernoulli}) shows that thermal driving is indeed negligible in the energy budget of the outflow, confirming that the solution is a cold MHD wind, not a magneto-thermal wind. It also shows that $\mathcal{B}>0$ indicating that the wind is free to escape the gravitational potential well up to infinity.

\begin{figure}[h!]
 	\centering
   \includegraphics[width=0.95\linewidth]{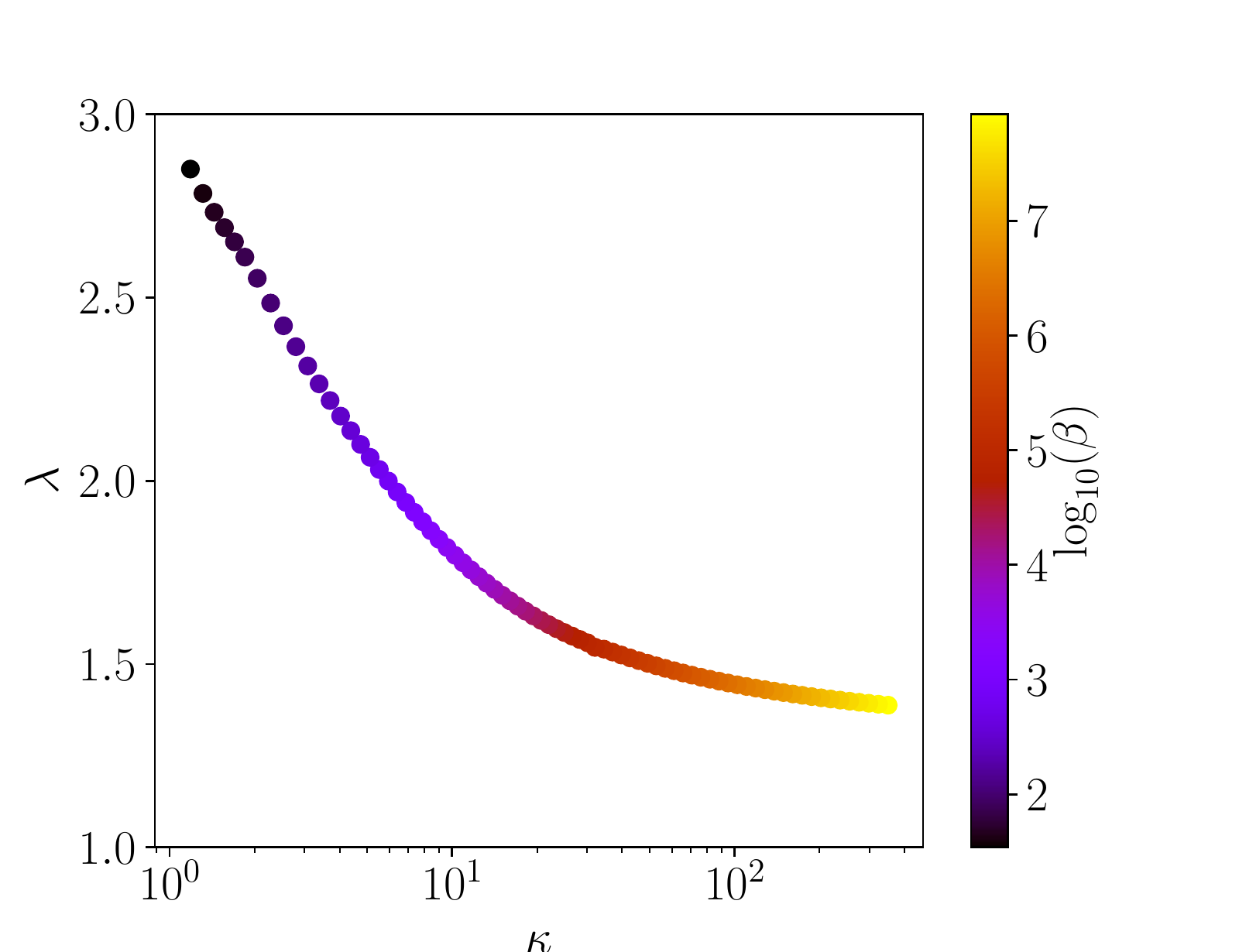}

      \caption{Dependence on $\kappa$ and $\lambda$ of the fiducial solutions as a function of the disc magnetisation $\beta$. As the field strength increases, $\lambda$ increases while the mass loading parameter decreases. }
         \label{fig:kappa-lambda}
 \end{figure}

 A systematic exploration of the $\kappa,\lambda$ parameters as a function of $\beta$ in the fiducial simulation gives Fig.~\ref{fig:kappa-lambda}. This figure can directly be compared to figure~2 in \cite{blandford82}. It is interesting to note that these new solutions at low magnetisation have a much larger $\kappa$ and much smaller $\lambda$ than the historical solutions of \cite{blandford82}. Such a trend was already observed by \cite{Jacquemin19} in ideal MHD for $\beta>1$. The solutions I obtain here have even smaller $\lambda$, with an asymptote in the high $\beta$ limit close to $\lambda=1.4$. This is smaller than the $\lambda=3/2$ limit demonstrated by \cite{blandford82}. It should be however pointed out that this limit is computed for a wind emitted from $z=0$, while here the wind is emitted from $z=5H$, so the lever arm is physically allowed to be slightly smaller than $3/2$. This is confirmed by analysing the Bernoulli invariant at $\beta=10^7$, which still shows that the magnetic energy is dominating the energetics at the disc surface (not shown).

\section{Parameter space exploration\label{sec:parameters}}
\subsection{Ambipolar diffusion}

\begin{figure*}[h!]
   \centering
   \includegraphics[width=0.95\linewidth]{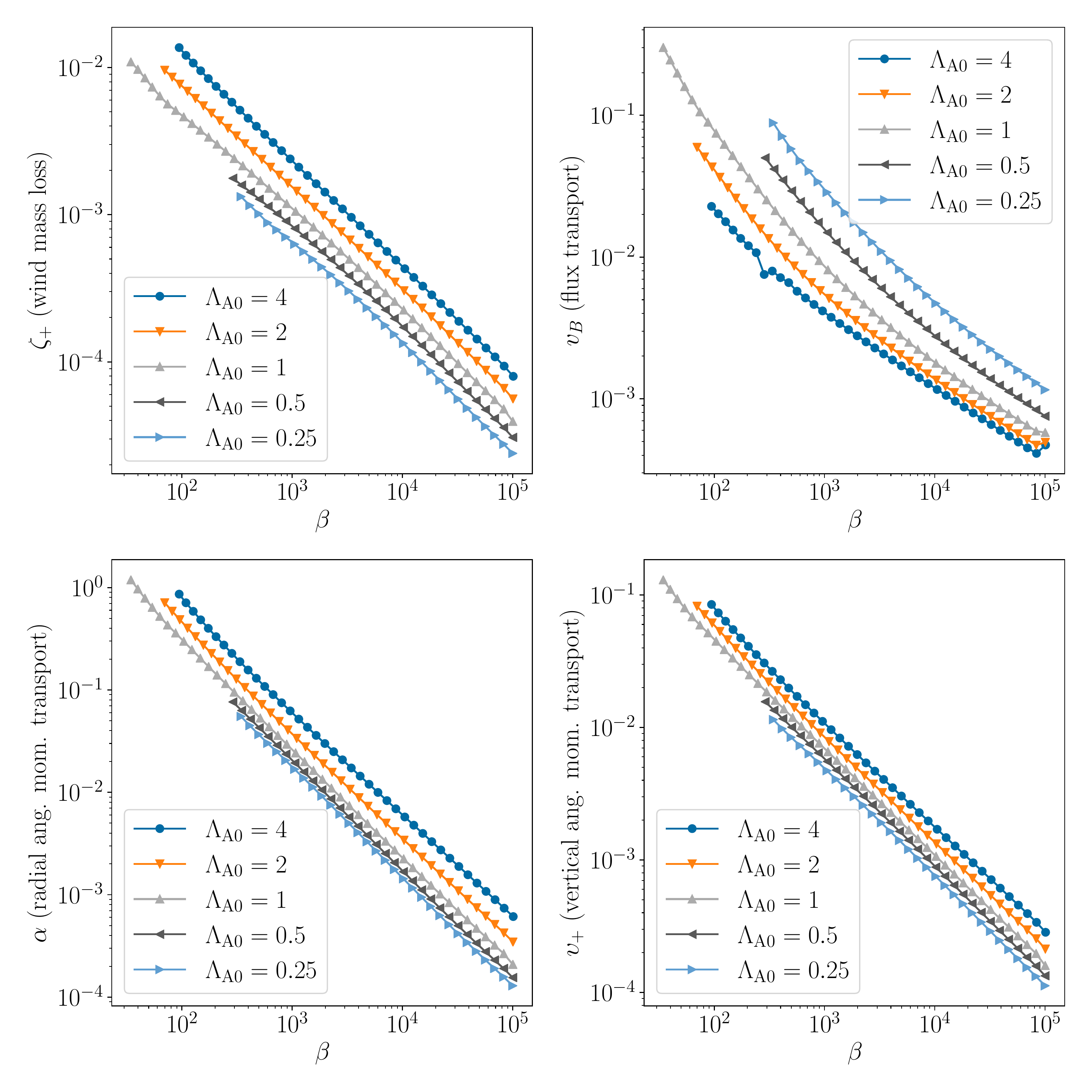}

      \caption{Transport coefficients as a function of the disc magnetisation and mid-plane Elsasser number ${\LA}_0$.}
         \label{fig:statAm}
   \end{figure*}

In order to explore the impact of the disk diffusivity on the behaviour of the outflow, I have varied the strength of ambipolar diffusion in the disc plane ${\LA}_0$ (eq.~\ref{LAdef}), keeping the same vertical profile. The resulting transport properties are shown in figure.~\ref{fig:statAm}. The general trend is that as diffusion is increased, the mass loss rate and angular momentum fluxes decrease, while magnetic flux transport increases. This overall suggest that wind-driven accretion tends to be reduced when ambipolar diffusion gets stronger, as expected naively. It should however be noted that all transport coefficients do not vary in the same proportion. Taking $\beta=10^4$ as the reference case, I find that a decrease of ${\LA}_0$ by a factor 16 leads to a decrease of $\zeta$, $\alpha$ and $\upsilon$ by respectively a factor $3.2$, $4.0$ and $2.25$, while $v_\psi$ increases by a factor $4.0$. This therefore suggests a relatively weak sensitivity of the transport coefficient on ${\LA}_0$ (shallower than $\LA^{0.5}$), especially for the vertical angular momentum coefficient $\upsilon$. The weak dependence of $\upsilon$ on the disc dissipation properties was already observed in shearing box simulations (see \citealt{lesur20}, figure 39 and related text)
 and is here confirmed in global geometry. Overall this confirms that wind-driven angular momentum extraction is only moderately affected by the strength of ambipolar diffusion, and that the mass accretion rate (\ref{eq:mdotfit}) is approximately valid (up to a factor of a few) for the range of ambipolar diffusion considered here.

\subsection{Ohmic diffusion}

In addition to the strength of ambipolar diffusion, I have also explored the impact of Ohmic diffusion on the flow topology. In protoplanetary discs, this leads to wind solutions valid in \cite{Gammie96} historical ``dead zone'', found around 1 AU in typical chemical models. To limit the computation time required by such models, I have chosen to limit myself to ${\Rm}_0>1$. In the literature, one often finds Ohmic diffusion evaluated in terms of Ohmic Elsasser number $\Lambda_O$, which depends on the field strength. One typically has $\Lambda_0={\Rm}_0\beta$ so ${\Rm}_0\gtrsim 1$ corresponds typically to the mid-plane value quoted at $R\lesssim 1\,\mathrm{AU}$ by several authors (\citealt{Bai13a}, figure~1; \citealt{Thi19}, figure~8). I should finally point out that in all of these models, I keep ambipolar diffusion fixed to its fiducial value ${\LA}_0=1$. Hence, Ohmic diffusion is \emph{added} to the diffusivity tensor of the fiducial run.
 
\begin{figure*}[h!]
   \centering
   \includegraphics[width=0.95\linewidth]{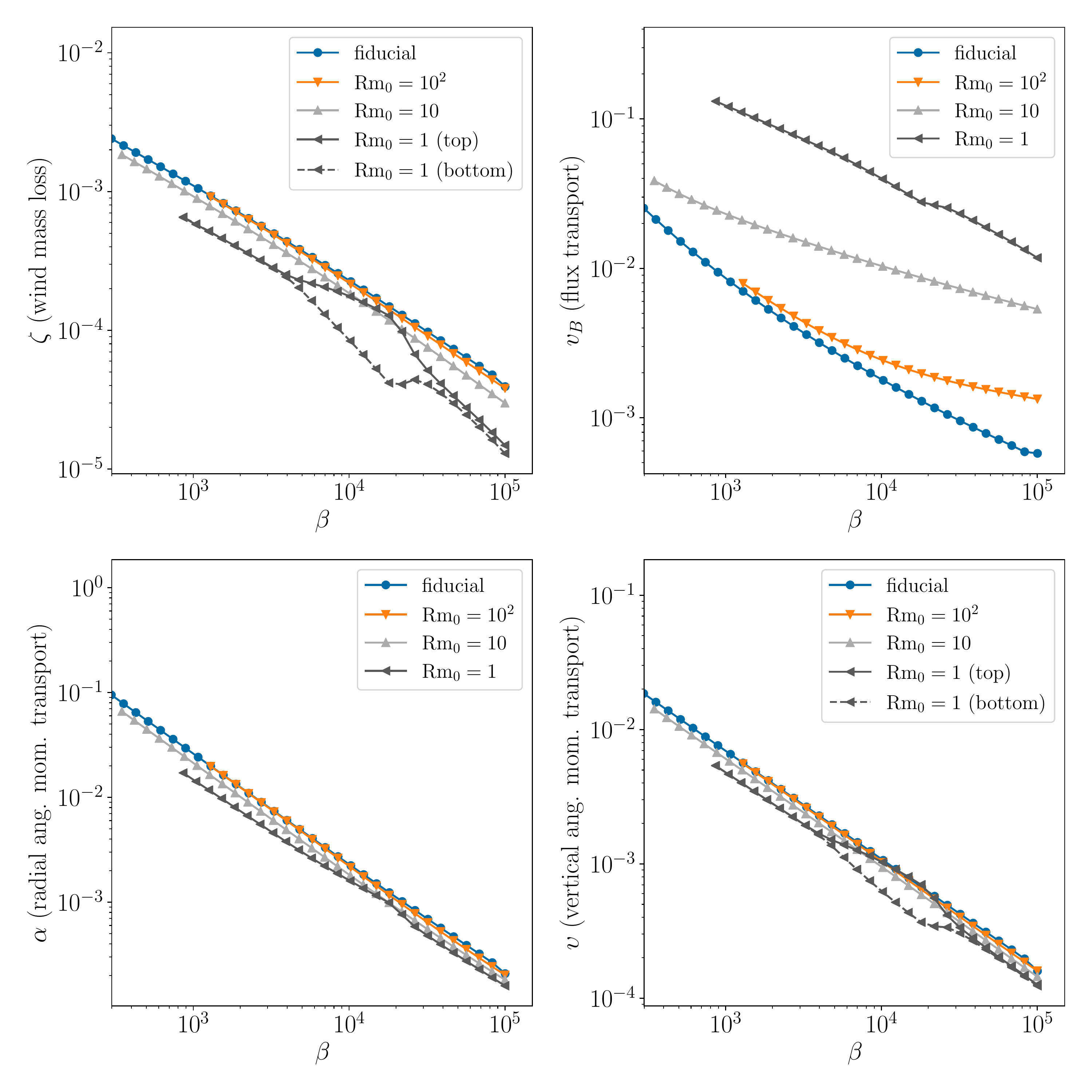}

      \caption{Transport coefficients as a function of the disc magnetisation and mid-plane Reynolds number ${\Rm}_0$. Note that these solutions also include ambipolar diffusion with ${\LA}_0=1$. The fiducial run is physically equivalent to ${\Rm}_0=\infty$.}
         \label{fig:statRm}
   \end{figure*}

	As for ambipolar diffusion, let me start with the transport coefficients as a function of ${\Rm}_0$(fig.~\ref{fig:statRm}). Here, the angular momentum transport coefficients $\alpha$ and $\upsilon$ are barely affected by Ohmic diffusion, even when ${\Rm}_0=1$. This suggests that the accretion rate in \cite{Gammie96} dead zone should not be so different from that observed in regions dominated by ambipolar diffusion. I note however that field diffusion rate is increased by possibly several orders of magnitude. This is probably the most stringent effect of adding Ohmic diffusion to the system. While ${\Rm}_0\geq 10$ solutions are top-down symmetric (so only $\zeta_+$ and $\upsilon_+$ are plotted), the ${\Rm}_0=1$ solution breaks this top-down symmetry around $\beta\simeq 10^4$. The direction of this symmetry breaking is random. Starting from slightly different initial condition, I could get $\zeta_+>\zeta_-$ or the opposite, so there is \emph{no physically preferred} direction for this symmetry breaking.
	
\begin{figure*}
   \centering
   \includegraphics[width=0.32\linewidth]{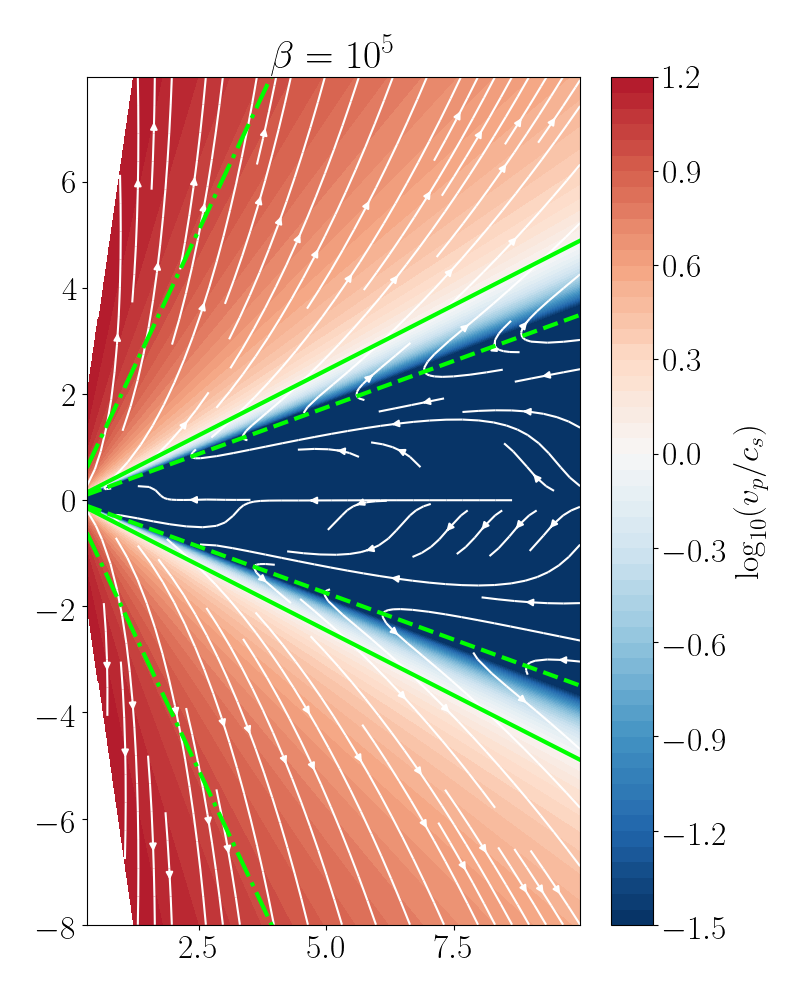}
   \includegraphics[width=0.32\linewidth]{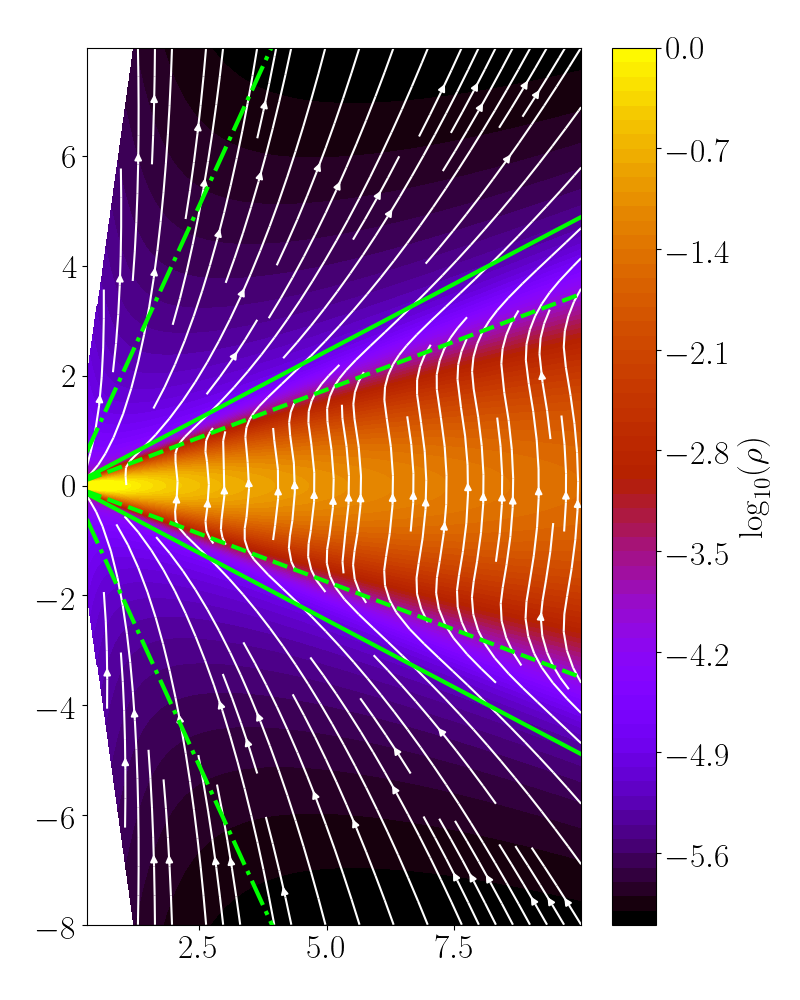}
   \includegraphics[width=0.32\linewidth]{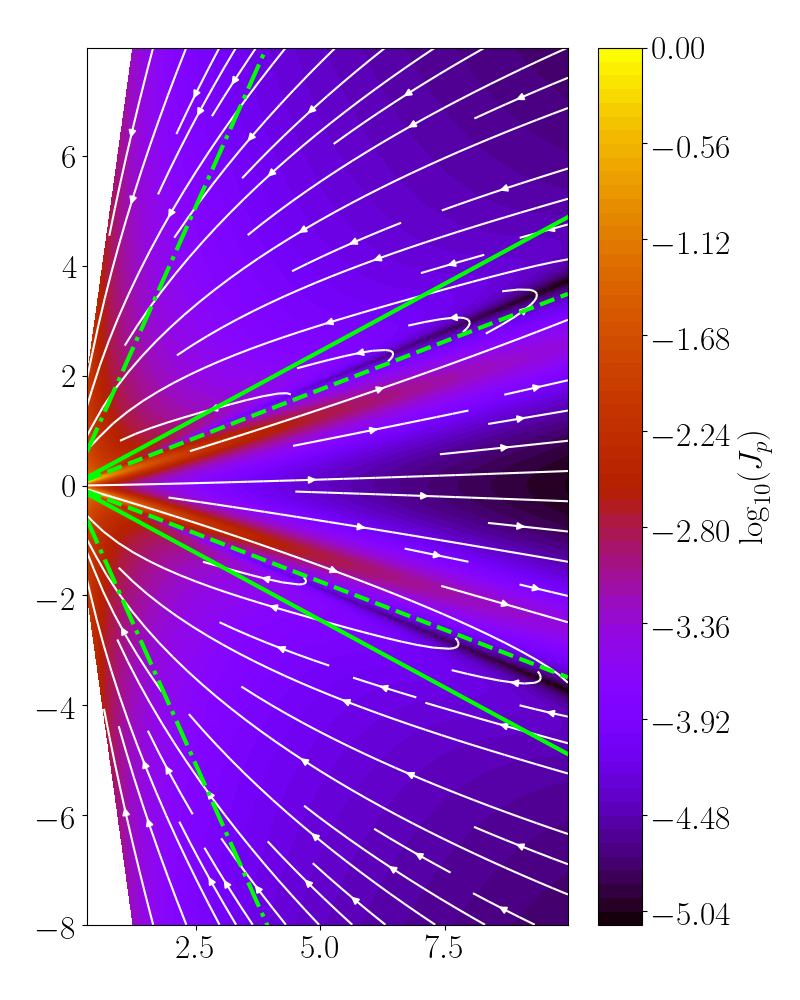}
   \caption{Flow structure for the solution ${\Rm}_0=1$ at $\beta=10^5$. Left: streamlines and sonic Mach number. Middle: field lines and density map. Right: Poloidal current lines and current density. Compared to the fiducial case, the current layer is now localised at the disc surface, leading to a surface accretion flow.\label{fig:flowRm1}}
\end{figure*}

\begin{figure}[h!]
   \centering
   \includegraphics[width=0.95\linewidth]{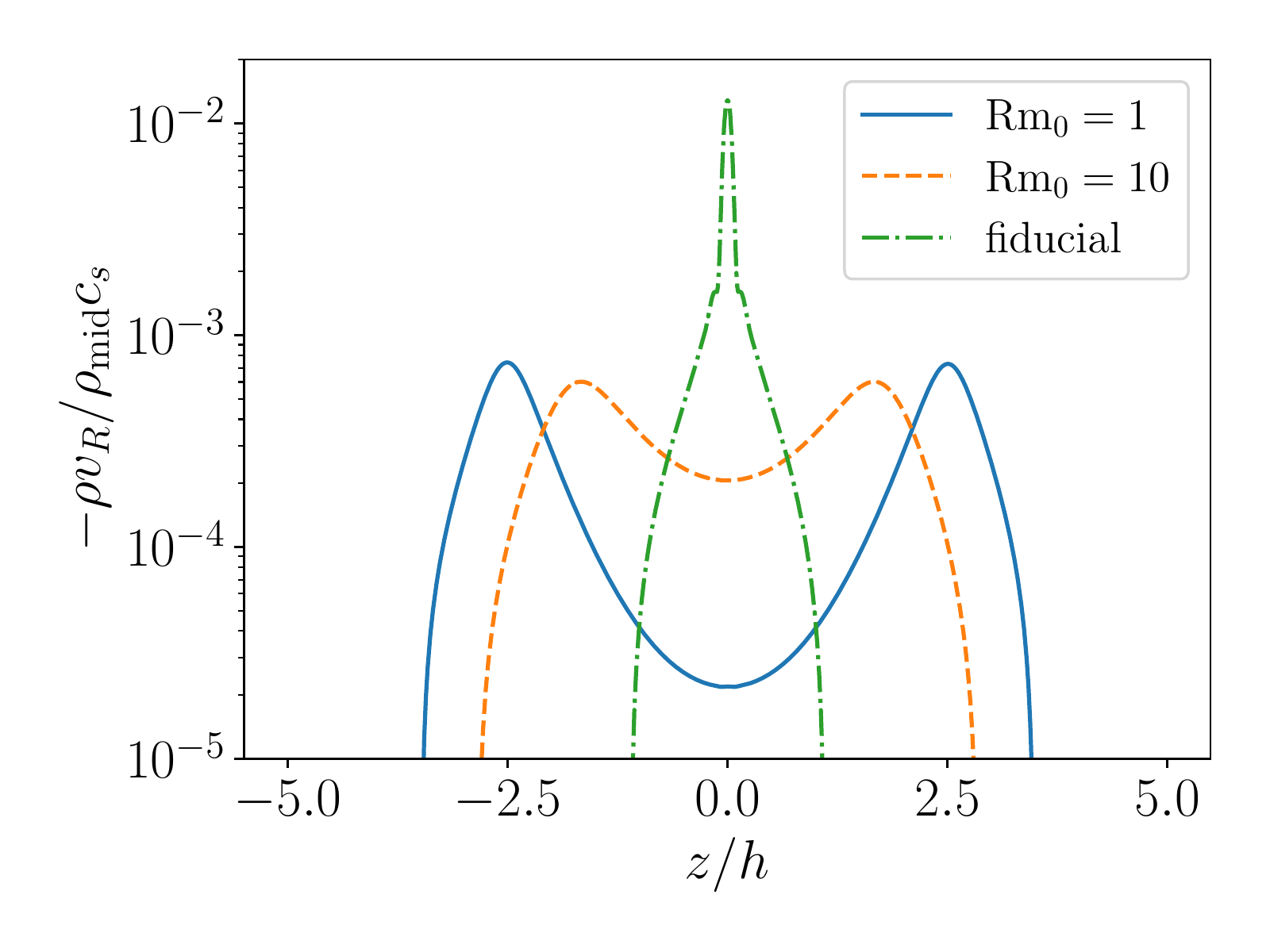}
   \includegraphics[width=0.95\linewidth]{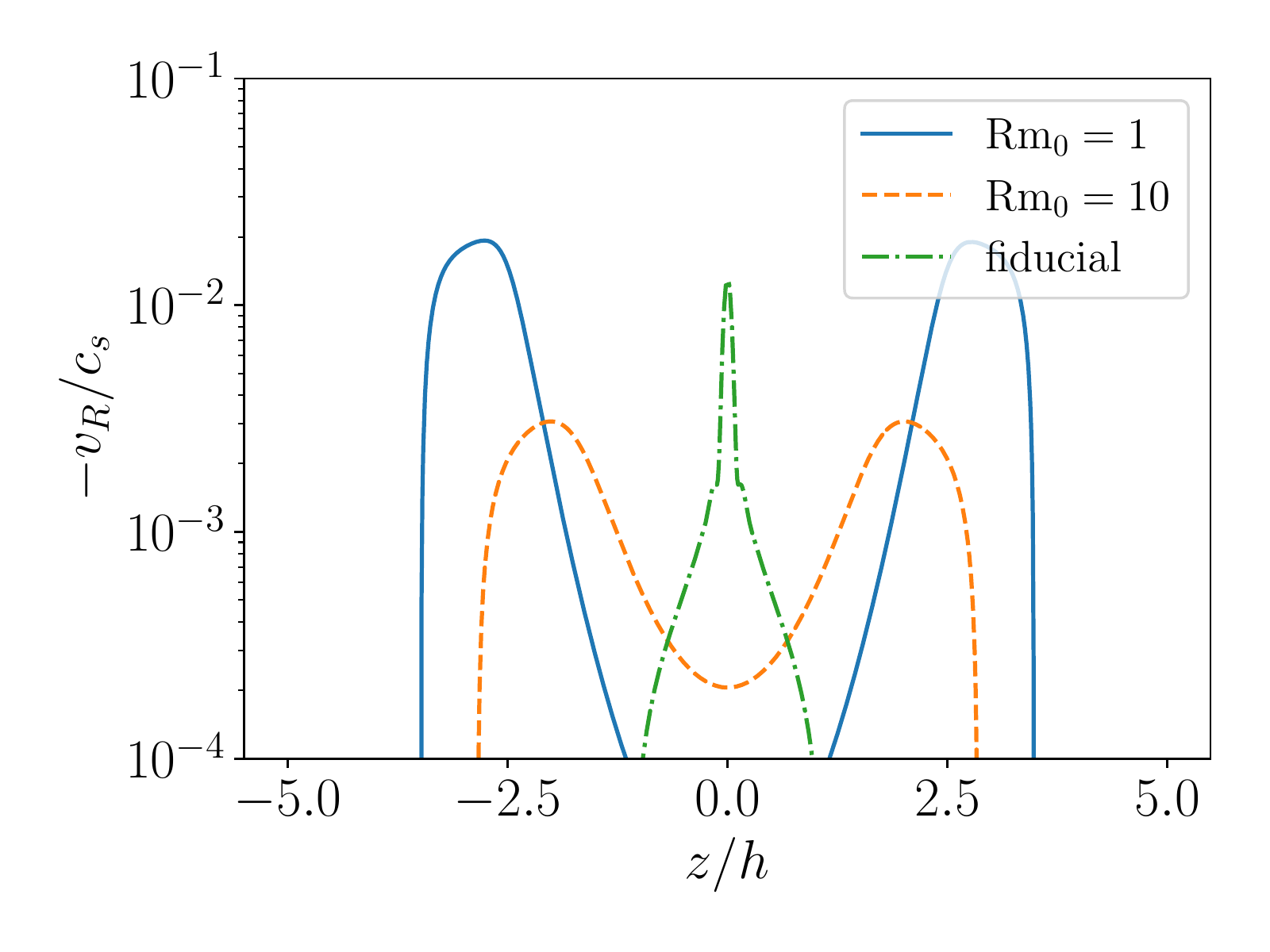}
   \caption{Inward mass flux (top) and radial velocity Mach number (bottom) as a function of Ohmic diffusion strength at $\beta=10^5$. Note that the accretion flow moves towards the disc surface as Ohmic diffusion is increased, following the poloidal current sheet (see text).\label{fig:accretionRm1}}
\end{figure}

	The fact that the transport coefficients are approximately unaffected by Ohmic diffusion should not lead to the wrong impression that the solutions are physically identical. Indeed, a careful examination of the solution at ${\Rm}_0=1$ (Fig.~\ref{fig:flowRm1}) shows that the solution in the disc is very different from the fiducial solution. First, the current sheet which was initially localised in the mid-plane (Fig.~\ref{fig:current}) is now localised on the disc surfaces. This can be easily interpreted as a result of Ohmic diffusion. Indeed, while ambipolar diffusion can have a focusing effect, Ohmic diffusion is really only a diffusive process. Hence, the current sheet cannot form in the strongly diffusive mid-plane, and is therefore pushed away, at the disc surface. This change of localisation of the current sheet in turn impacts the accretion flow, since accretion is driven by the magnetic torque created by the poloidal current (see section \ref{sec:current}). This is more evidently seen in Fig.~\ref{fig:accretionRm1}, where the accretion flow clearly moves from the disc mid-plane towards the disc surface as ${\Rm}_0$ is decreased, thereby following the localisation of the current layer. In the end, poloidal current and accretion flows are both localised at the disc surface. However, since the current is set globally by the electrical circuit enforced by the wind, its intensity is not really affected by this change in localisation, implying that total torques and accretion rates are not dramatically modified, as indicated by transport coefficients.
	
  \begin{figure*}
   \centering
   \includegraphics[width=0.32\linewidth]{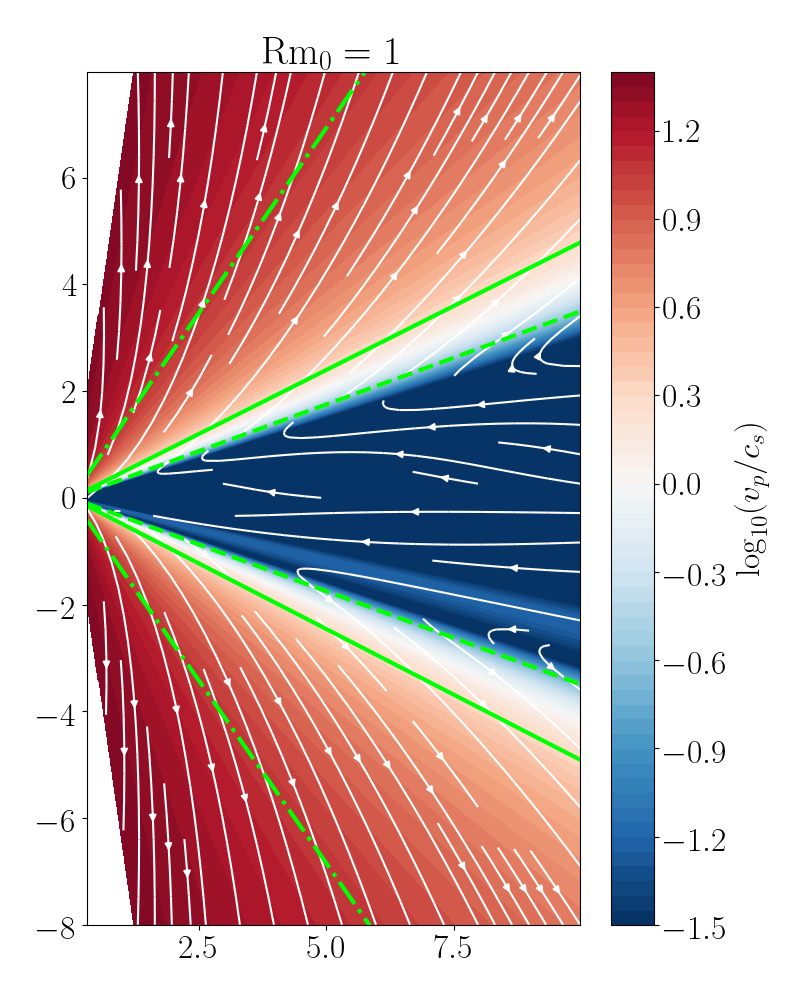}
   \includegraphics[width=0.32\linewidth]{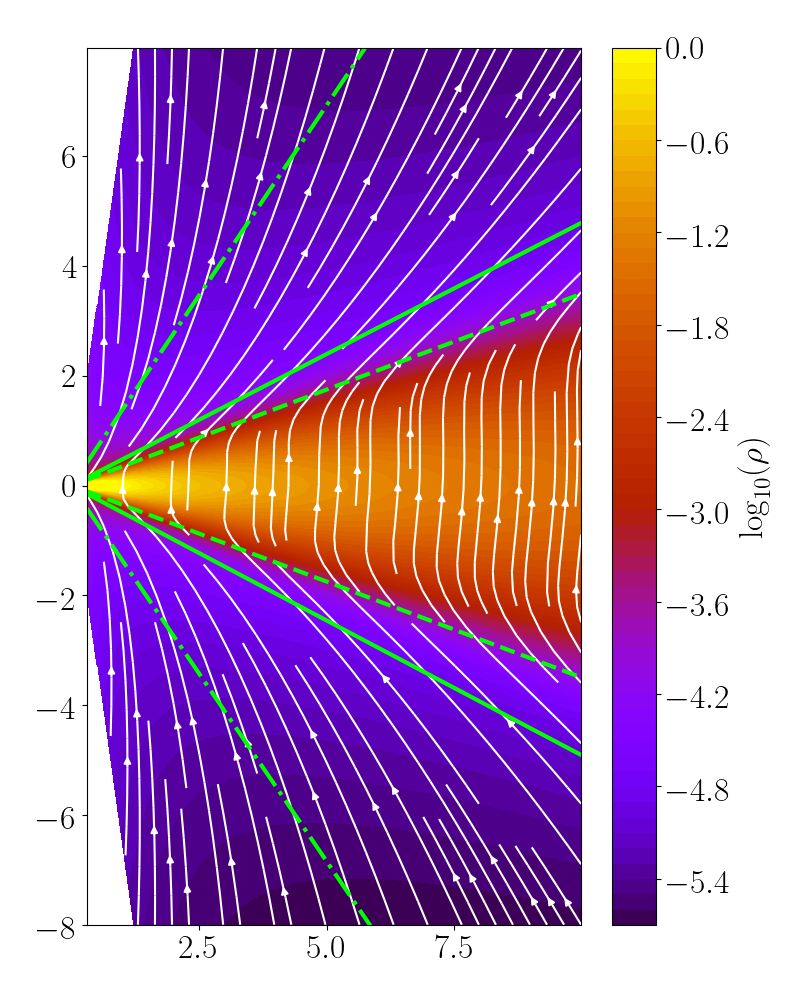}
   \includegraphics[width=0.32\linewidth]{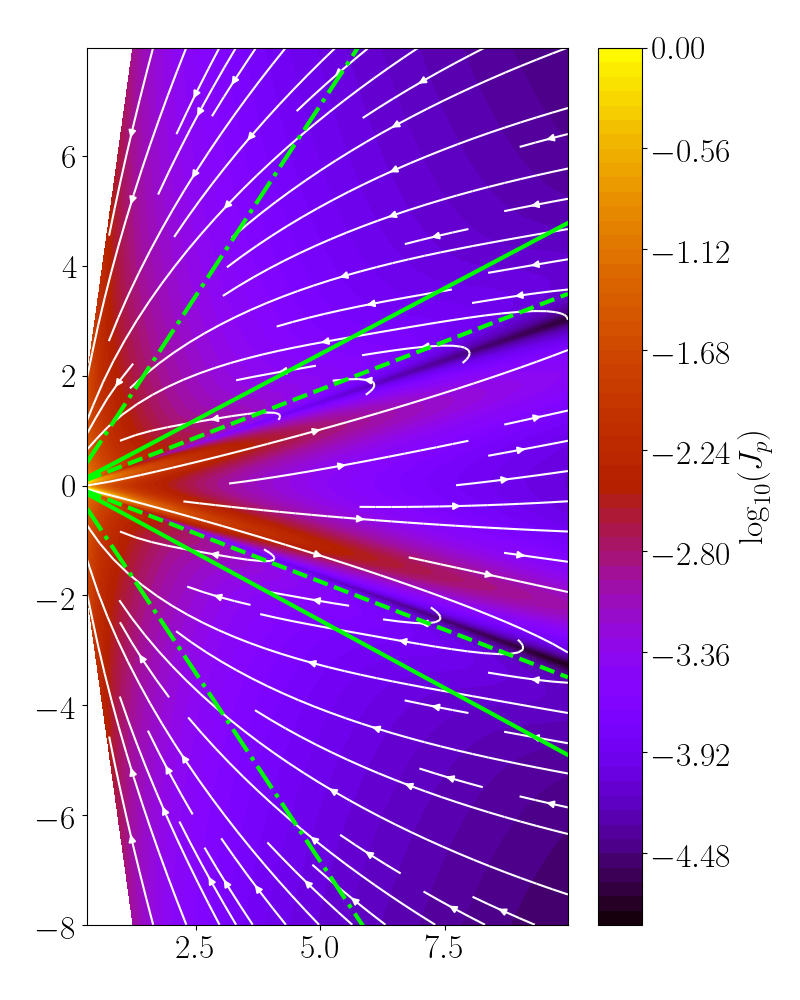}
   \caption{Flow structure for the solution ${\Rm}_0=1$ at $\beta=10^4$. Left: streamlines and sonic Mach number. Middle: field lines and density map. Right: Poloidal current lines and current density. \label{fig:flowDissymetric}. This solution exemplify the dissymmetric solutions found around $\beta=10^4$ at ${\Rm}_0=1$}
\end{figure*}

Having understood that the current sheet gets divided into two parts and pushed towards the surface, one can easily imagine that such a process might not be totally symmetric for some range of parameters. This is what happens in the ${\Rm}_0=1$ solution around $\beta=10^4$. In this case, the current sheet is more pronounced in the southern hemisphere (Fig.~\ref{fig:flowDissymetric}), leading to an accretion stream more pronounced on the bottom side of the disc, and inclined field lines in the disc bulk. I should stress that while accretion occurs in the southern side of the disc in this particular example, the wind is more pronounced on the northern side of the disc since $|\zeta_+|>|\zeta_-|$ and $\upsilon_+>\upsilon_-$. Hence angular momentum and mass flow away on the side \emph{opposite} to that of accretion. Note that such a behaviour was also observed by \cite{Bethune17} in dissymmetric solutions.


\section{Discussion and Conclusions}
In this work, I have used a finite volume code to derive self-similar wind solutions applicable to the diffusive regime of protoplanetary discs. In contrast to previous self-similar solutions published to date, no assumption is made on flux stationarity or top/down symmetry. It also circumvent the difficulties of full-blown 3D numerical simulations which are often limited by their integration time and inner boundary conditions. Using this technic, it is possible to explore systematically a wide range of parameters at a reduced numerical cost.

Using this tool, I have presented a series of wind solutions ranging from $\beta=10^{8}$ to $\beta=35$ valid in the non-ideal regions of protoplanetary discs $R\gtrsim 1\,\mathrm{AU}$. I have shown that cold wind solutions (i.e. that do not require atmospheric heating) exist in this entire range of parameters. Some of these solutions are very similar to the one found using 2.5D and 3D simulations, in the same range of parameters. These solutions exhibit several important properties, some of which can be confirmed observationally.

First and foremost, magnetised disc winds \emph{always exist} and are \emph{unavoidable} as soon as a large scale magnetic field threads the disc. This statement is true even for $\beta=10^{8}$ fields which correspond to a few $\mu\mathrm{G}$ at 10 AU for typical surface densities. They give accretion rates $\dot{M}_\mathrm{acc}$ (eq.~\ref{eq:mdotB}) which depends essentially on the field strength, and which are broadly speaking compatible with observed accretion rates onto T-Tauri stars provided that $10^3<\beta<10^5$ or equivalently $B_z\sim$ a few mG. Interestingly, $\dot{M}_\mathrm{acc}$ depends only weakly on the disc surface density, in contrast to viscous disc models. However, the direct comparison between the accretion rate in the disc bulk (as measured by eq.~\ref{eq:mdotB}) and the mass accretion measured \emph{onto the star} is probably misleading. The latter is probably significantly smaller than the former because the ejection index (see below) is of order one, hence the \emph{disc} mass accretion rates could be significantly larger (possibly by an order of magnitude) than the ones quoted here.

The ejection index $\xi$, which essentially quantifies the ratio of ejected to accreted mass, depends only weakly on the magnetic field strength and is of the order of (but slightly less than) unity for the range of magnetisation I have explored. This means that \emph{the mass loss rate is comparable to the mass accretion rate} at a given radius. Consequently, if one assumes steady state, the accretion rate is expected to increase significantly with radius. As expected from a high $\xi$ wind \citep[eq. 11.22]{lesur20}, the wind lever arm is always relatively small with typically $\lambda<2$ for $\beta>10^4$, without requiring any heating. Such a high $\xi$ and low $\lambda$ was already found in previous numerical work \citep{Bethune17,Bai17}, and it was initially thought that atmospheric heating, included in all these models, was the main reason for this result, following the argument of \cite{Casse00b}. I show here that heating is actually not key to get these high $\xi$-low $\lambda$ solutions, a result also reported recently by \cite{Jacquemin19} in the context of fully ionised discs. Indeed, the solutions presented here are locally isothermal and I have shown that thermal heating is negligible in the wind energy budget even for $\beta=10^{8}$. Hence, low $\lambda$ and high $\xi$ are quite clearly signatures of weakly magnetised ($\beta\gg 1$) outflows.

The vertical field strength predicted to get accretion rates compatible to typical T-Tauri stars (about a mG at 10 AU, see eq.~\ref{eq:mdotB}) are compatible with recent upper limits obtained from Zeeman measurements \citep[e.g.][]{Vlemmings19}. When it comes to field detection, it is quite clear that the toroidal component is a better candidate than the vertical one since the former is expected to be much stronger (by a factor 10 or so at $\beta\sim 10^4$) than the latter (eq.~\ref{eq:magshear}). Note however that the toroidal component is expected to change sign across the mid-plane, so this detection is possible only for tracers which do not average out the field through the disc thickness.

When it comes to the disc microphysics, I find that the disc-averaged mass accretion \& ejection rates depend only weakly on ambipolar and Ohmic diffusivities. This implies that one does not need very elaborated ionisation models in the disc to get the right accretion/ejection bulk properties. However, the disc vertical structure actually strongly depends on diffusion. Solutions with large Ohmic diffusion (i.e. valid closer to 1 AU) tend to exhibit accretion at the disc \emph{surface} while weaker Ohmic diffusion ($R>10\,\mathrm{AU}$) show accretion in the disc mid-plane. Some of the strong Ohmic diffusion solutions also exhibit top/down dissymmetry, where accretion mostly occurs mostly on one side of the disc. Such a dissymmetry was also found in simulations including Ohmic and ambipolar diffusion \citep{Bethune17,Gressel20} in similar range of parameters, so this result is not unheard of. These topology properties will likely have a strong impact on dust growth and planet migration theory, but again, they barely affect the vertically-averaged transport properties.

Finally, In all of these models, the large scale magnetic field is found to be transported \emph{outwards}, in agreement with \cite{BaiStone17}. However, the transport velocity is about an order of magnitude lower than that measured by \cite{BaiStone17} in similar conditions (ambipolar diffusion only). Moreover, I find that the transport rate varies significantly with the strength of Ohmic diffusion and to a lesser extent with Ambipolar diffusion. Overall,  This dependency might explain why simulations and models tend to disagree on the magnetic transport rate \citep{BaiStone17,leung19,Gressel20}. 

Clearly, the large scale field transport is the main bottleneck of a complete theory of wind-driven accretion since the strength of the large scale poloidal field is the main control parameter. This cannot be ignored for the secular evolution of discs since the field strength is expected to vary with time. Indeed, while class II objects seem to require field strengths of a few mG at 10s of AU to get the right accretion rates (see above), core collapse calculations tend to suggest  a field strength of 100 mG when the disc forms \citep[e.g.][]{Masson16}. Hence, the field strength must be reduced by at least one order of magnitude (more probably two) between class 0 and class 2, indicating that flux transport must be relatively efficient. The problem of field transport should therefore deserve more attention in the future if winds are to be used in secular evolution models.

\begin{acknowledgements}
      I am thankful to Jonatan Jacquemin-Ide, Etienne Martel, Jonathan Ferreira and Cornelis Dullemond for fruitful discussions regarding the physics of self-similar solutions. I acknowledge support from the European Research Council (ERC) under the European Union Horizon 2020 research and innovation programme (Grant agreement No. 815559 (MHDiscs)). All of the computations presented in this paper were performed using the GRICAD infrastructure (https://gricad.univ-grenoble-alpes.fr), which is supported by Grenoble research communities.
\end{acknowledgements}

%
%

\bibliographystyle{aa}
\bibliography{biblio}
\end{document}